\documentclass[aps,preprint]{revtex4}
\usepackage{amsfonts}
\usepackage{mathrsfs}
\usepackage{graphicx}
\usepackage{subfigure}
\usepackage{amsmath}
\usepackage{amssymb}
\usepackage{amsbsy}
\usepackage{bm}

\usepackage{algorithm}
\usepackage{algorithmicx}
\usepackage{setspace}

\begin{document}

\title{Statistical mechanics of continual learning: variational principle and mean-field potential}
\author{Chan Li$^{1}$}
\thanks{Equal contribution.}
\author{Zhenye Huang$^{2}$}
\thanks{Equal contribution.}
\author{Wenxuan Zou$^{1}$}
\thanks{Equal contribution.}
\author{Haiping Huang$^{1}$}
\email{huanghp7@mail.sysu.edu.cn}
\affiliation{$^{1}$PMI Lab, School of Physics,
Sun Yat-sen University, Guangzhou 510275, People's Republic of China}
\affiliation{$^{2}$CAS Key Laboratory for Theoretical Physics, Institute of Theoretical Physics,Chinese Academy of Sciences, Beijing 100190, People's Republic of China}

\date{\today}

\begin{abstract}
An obstacle to artificial general intelligence is set by continual learning of multiple tasks of different nature.
Recently, various heuristic tricks, both from machine learning and from neuroscience angles, were proposed, but they 
lack a unified theory ground. Here, we focus on continual learning in single-layered and multi-layered neural networks
of binary weights. A variational Bayesian learning setting is thus proposed, where the neural networks are trained in a field-space,
rather than gradient-ill-defined discrete-weight space, and furthermore, weight uncertainty is naturally incorporated, and
modulates synaptic resources among tasks. From a physics perspective, we translate the variational continual learning into 
Franz-Parisi thermodynamic potential framework, where previous task knowledge acts as a prior and a reference as well.
We thus interpret the continual learning of the binary perceptron in a teacher-student setting as a Franz-Parisi potential computation.
The learning performance can then be analytically studied with mean-field order parameters, whose predictions coincide with 
numerical experiments using stochastic gradient descent methods. Based on the variational principle and Gaussian field approximation of internal preactivations in hidden layers,
we also derive the learning algorithm considering weight uncertainty, which solves the continual learning with binary weights using multi-layered neural networks,
and performs better than the currently available metaplasticity algorithm where binary synapses bear hidden continuous states and the synaptic plasticity is modulated by a heuristic 
regularization function.
Our proposed principled frameworks also connect to elastic weight consolidation, weight-uncertainty modulated learning,
and neuroscience inspired metaplasticity, providing a theory-grounded method for the real-world multi-task learning with deep networks.
\end{abstract}

 \maketitle

\section{Introduction}
The environment an intelligent agent faces is commonly highly structured, and moreover, multiple tasks encoding this structure occur in sequence.
Therefore, it is important for the agent to learn the continually evolving structures embedded in sequential tasks, i.e., transfer the knowledge gained from previous experiences to
the learning of a current novel or unfamiliar task. However, during this continual learning, it is well-known that the previous task knowledge may be erased after learning a new task (so-called
catastrophic forgetting~\cite{CF-1989,Rev-2019}). Uncovering neural mechanisms underlying a successful continual learning especially in the natural world
presents a challenge in the current AI and even neuroscience research. There also emerge recently interesting works on biological neuronal networks in this regard~\cite{Cell-2019,Yang-2021,Saxe-2022}, and the neuroscience research provides in turn 
insights for improving the performance of continual learning in artificial neural networks~\cite{PNAS-2018,NC-2020,meta-2021}.

To avoid catastrophic forgetting, the machine learning community also proposed many heuristic strategies. For example, the elastic weight consolidation method introduces the Fisher information matrix 
to measure weight importance in the consecutive task learning~\cite{PNAS-2017}, which is further improved by tracking individual weight contribution over the entire dynamics of training loss~\cite{Zenke-2017}.
An attention mask can also be learned to alleviate the catastrophic forgetting~\cite{Att-2018}. Another important line is using the Bayesian approach~\cite{Gal-2019}. This line shows that
the synaptic uncertainty plays a significant role in taking the learning trade-off between two consecutive tasks~\cite{TaskAg-2018,VCL-2018,VCL-2020,CLAW-2020}. We remark that these heuristic strategies have diverse design principles,
but from a statistical physics perspective, they can be put under a unified framework of variational mean-field theory. Although recent theoretical works focused on phase transitions in transfer learning from source task to target task~\cite{Lu-2021}
and on-line learning dynamics of teacher-student setup~\cite{Lee-2019,JPSJ-2021,Seba-2021}, these works did not take into account weight uncertainty, which is an essential factor in learning neural networks~\cite{Prob-2013}, including more efficient and robust binary-weight networks.
In addition, a recent study pointed out that the concept of meta-plasticity from brain science plays a key role in the continual learning of binary-weight neural networks~\cite{meta-2021}.
This concept highlights that the binary synapse bears a hidden continuous state, and the synaptic plasticity is modulated by a heuristic regularization function.
Our theoretical framework demonstrates that a variational principle can be constructed to explain the role of synaptic uncertainty, and moreover, the knowledge-transfer between tasks can be actually captured by
a thermodynamic potential~\cite{Huang-2014}, from which the learning performance can be predicted. 

In this work, we not only carry out a thorough theoretical analysis of a toy teacher-student learning setting, where both tasks of a certain level of similarity are learned in sequence, but also 
apply the same principle to deep continual learning of structured datasets, which demonstrates the effectiveness of the variational mean-field principle, especially in the binary-weight neural networks
where only meta-plasticity was previously proposed. Overall, our theory bridges statistical physics, especially the concept of the Franz-Parisi
potential, originally studied in mean-field spin glass models~\cite{Franz-1995}, to theoretical underpinnings of the challenging continual learning. This connection may prove fruitful in future researches.

\section{Continual learning with binary perceptron}
The binary perceptron offers an ideal candidate for understanding non-convex learning, as a theoretical analysis is possible by using statistical physics methods~\cite{HH-2022}. 
	Here, we will use a teacher-student setting to perform the theoretical analysis of variational
	continual learning,
	in which the ground truth network is quenched before learning. In this section of toy model analysis, we use $\boldsymbol\xi$ and $\boldsymbol{W}$ to indicate
	the student's and teacher's weights, respectively. In the next section of training deep networks (no ground truth in this case), we use $\mathbf{w}$ to indicate the weights to learn.
\subsection{Learning setting}
The standard perceptron is a single-layered network with $N$ binary input nodes, $x_i = \pm1$ $(i=1,2,...,N)$, 
	and a single binary output node, $y = \pm1$, which is connected by $N$ binary weights $\xi_i = \pm1\, (i = 1,2,...,N)$. 
	Given an input $\bm{x}$, the output is specified by $y = \mathrm{sign}(\frac{1}{\sqrt{N}}\sum_ix_i\xi_i)$, where $\mathrm{sign}(x)$ is the sign 
	function. A perceptron can be used to classify inputs according to their respective labels ($\pm1$ here).
	A statistical mechanics analysis revealed that the network can store up to a critical threshold of pattern density (or sample complexity) 
	$\alpha \simeq 0.83$~\cite{Krauth-1989},
	where $\alpha = \frac{M}{N}$ is the random-pattern (as inputs) density,
	and $M$ is the number of random patterns. Instead of this classic random pattern storage setting,
	we consider learning task of
	random patterns with respective labels generated by teacher networks (corresponding to different tasks).
	This is called the teacher-student setting~\cite{Gyo-1990,Sompolinsky-1990}, where the student network learns to infer the teachers' rule embedded in the supplied data.
	
	With increasing number of supplied learning examples, the size of the candidate-solution space of weights shrinks, and thus the generalization error on fresh data examples decreases.
	The statistical mechanics analysis also predicted that at $\alpha \simeq 1.245$, a first order phase transition to perfect generalization occurs~\cite{Gyo-1990,Sompolinsky-1990}, which is
	the single-task learning.
	In our continual learning setting, we design two teacher networks with binary weights $\bm{W^1} \in \{\pm1\}^N$
	and $\bm{W^2} \in \{\pm1\}^N$, respectively. Both teacher networks are ground truth for corresponding tasks. 
	By definition, both teachers share an adjustable level of correlations in their weights, representing the similarity across tasks. In practice,
	their weights follow a joint distribution as
	\begin{equation}
	 P( \bm{W^1},\bm{W^2} ) = \prod_{i=1}^{N} P_0(W_{i}^1, W_{i }^2) = \prod_{i=1}^N \left( \frac{1+r_0}{4} \delta (W_{i}^1- W_{i }^2) + \frac{1-r_0}{4}\delta (W_{i}^1+W_{i }^2) \right),
	\end{equation}
	where $r_0 \in [-1,1]$ denotes the task similarity. $r_0$ also denotes the overlap of the two teacher
	networks, since $r_0=\frac{1}{N}\sum_{i=1}^NW^1_iW^2_i$. The marginal joint
	probability $P_0(W_{i}^1, W_{i }^2)$ can 
	be rewritten as $P_0(W_{i}^1, W_{i}^2) = p(W_{i}^1)p(W_{i}^2|W_{i}^1)$, 
	where $p(W_{i}^1) = \frac{1}{2}\delta (W_{i}^1 - 1) + \frac{1}{2}\delta (W_{i}^1 + 1)$,
	and $p(W_{i}^2|W_{i}^1) = \frac{1+r_0}{2} \delta (W_{i}^1 - W_{i}^2) + \frac{1-r_0}{2}\delta (W_{i}^1 + W_{i}^2)$. 
	To generate weights of the two teacher networks, we can first generate a set of random binary weights
	from the Rademacher
	distribution, and then flip the weight by a probability $\frac{1-r_0}{2}$.
	The random patterns for the two tasks are independently sampled
	from the Rademacher distribution as well, and then we have the training dataset $\{\bm{x}^{t,\mu}\}_{\mu=1}^{M_t}$, where the task index $t = 1,2$.
	Given the sampled patterns, the teacher networks generate corresponding labels for each task,
	$ \{y^{t,\mu} \}_{\mu = 1}^{M_t}$. 
	Hereafter, we use $\mathcal{\bm{D}}_t = \{\bm{x}^{t,\mu}, y^{t,\mu}\}_{\mu=1}^{M_t}$, to denote
	the two datasets corresponding to the consecutive two tasks.  
	The student network is another binary perceptron, whose goal is
	to learn the task rule provided by the teacher networks. 
	
	 In the above setting, the student network shares the same structure (connection topology)
	 with the two teacher networks, which implies that the
	student can not simultaneously learn both tasks perfectly, depending on the task similarity. 
	However, this setting allows us to explore how the student adapts its weights
	to avoid catastrophic forgetting during learning of a new task, and how the network takes a trade-off between 
	new and old knowledges in continual learning. 
	Studying this simple system could also provide us insights about the continual learning in more complex applications,
	such as deep learning in real-world data.
	
\subsection{Variational learning principle}
 Instead of training point weights, we consider
	learning the distribution of the weights in the sense that 
	we train the student network to find an optimal distribution
	of weights~\cite{Li-2020}.  Along this line, the variational method is an ideal framework for neural network learning~\cite{PRL-2018,Huang-2020},
	since we can use simple trial distribution to approximate the original intractable weight distribution.
	The learning becomes then finding an optimal trial distribution parameterized by variational parameters~\cite{HH-2022}. 	
	
	\begin{figure}
\centering
     \includegraphics[scale=1.0]{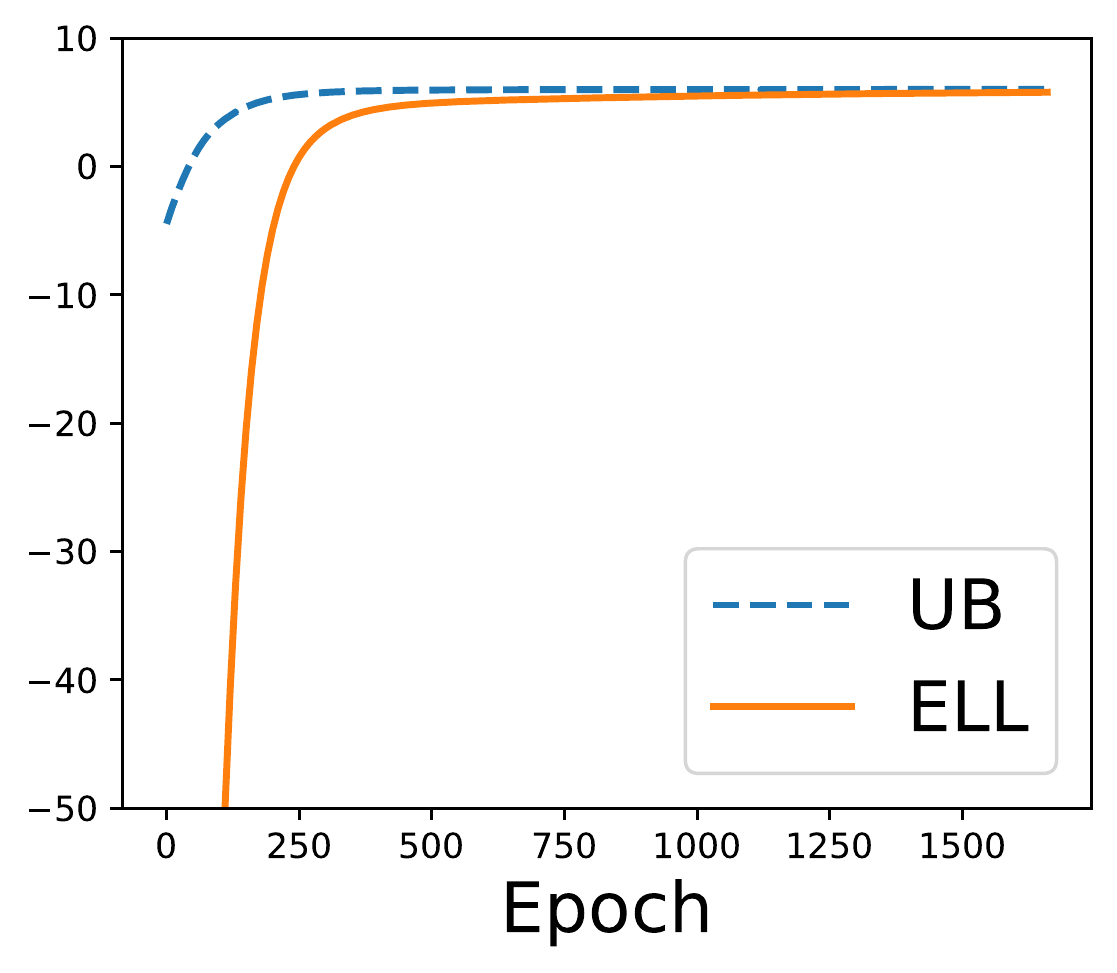}
  \caption{Comparison between expected log-likelihood (ELL) and its upper bound (UB) in a simple network with $10$ synapses and $M=10$ examples to learn.
  In this case, ELL can be exactly computed by an exhaustive enumeration. For the numerical purpose, we use the surrogate $\Theta(x)=\lim_{\kappa\to\infty}
  \frac{e^{\kappa x}}{2\cosh(\kappa x)}$. We take $\kappa=10.0$. Equation~\eqref{taskone} is used for ELL, while Eq.~\eqref{opt1} is used for UB.}
	\label{tight}
\end{figure}	

   In the first task, we introduce a variational distribution for synaptic weights, $
	q_{\bm{\theta}}(\bm{\xi}) = \prod_{i} \frac{e^{\beta\theta_i\xi_{i}}}{\cosh(\beta\theta_i)}$,
	where $\bm{\theta}$ are variational parameters, and $\beta$ is a hyperparameter. The optimal distribution can be approximated by maximizing the expected log-likelihood
	\begin{equation}\label{taskone}
		\begin{aligned} \bm{\theta} ^ { * }  = \arg \max_{ \bm{\theta} }   \mathbb{E}_{q_{\bm\theta}} \ln P(\mathcal{\bm{D}}_1|\bm{\xi}). \end{aligned}
	\end{equation}
	Given an input $\bm{x}$, we choose the probability $P(\mathcal{D}_1|\bm{\xi})=P(y | \bm{x},\bm{\xi})$ as $
	P(y | \bm{x},\bm{\xi}) = \Theta\Bigl(y\sum_{i=1}^{N}\xi_ix_i\Bigr)$,
	where $\Theta(x)$ is the Heaviside function such that $\Theta(x) = 1$, if $x>0$ and $\Theta(x) = 0$ otherwise. 
	In practice, based on the Jensen's inequality, we actually update the variational parameters by maximizing
	the upper bound  $\ln \mathbb{E}_{q_{\bm\theta}}  P(\mathcal{D}_1|\bm{\xi})$, which is less computationally challenging than the original one, and 
	the optimization problem can then be formulated as
	\begin{equation}\label{opt1}
		\begin{aligned}
		\bm{\theta} ^ { * } &= \arg \min _ { \bm{\theta} } \{- \ln  \mathbb{E}_{q_{\bm{\theta}}}P(\mathcal{\bm{D}}_1|\bm{\xi})\} \\
		&= \arg \min _ { \bm{\theta} } \Biggl\{- \ln  \mathbb{E}_{q_{\bm{\theta}}}\prod_{\mu}\Theta\Bigl(y^{\mu}\sum_{i=1}^{N}\xi_i x_i^{\mu}\Bigr ) \Biggr\}.
		\end{aligned}
	\end{equation}
	Maximizing the upper bound has been proved to be effective in unsupervised learning with many hidden neurons~\cite{Huang-2020}.
	When the number of weights are about $10$, the expected log-likelihood can be exactly computed, and 
	we have checked that the bound could be tight (see Fig.~\ref{tight}).
	Based on the assumption of large $N$ and the central limit theorem, Eq.~(\ref{opt1}) can be recast into the following form~\cite{PRL-2018}
	\begin{equation}
		\bm{\theta} ^ { * } = \arg \min _ { \bm{\theta} } \left\{- \sum_{\mu}\ln H\left(-\frac{y^{\mu}\sum_i x_i^{\mu}\tanh \beta\theta_{i}} { \sqrt{\sum_{i}(1 - \tanh^2 \beta\theta_i)} }\right) \right\},
	\end{equation}
	where $H(x) = \int_x^{\infty} dz e^{-\frac{z^2}{2}} / \sqrt{2\pi}=\int_x^{\infty}\mathcal{D}z$, where $\mathcal{D}z$ denotes a standard Gaussian measure. 
	Setting the loss function as $\mathcal{L} = - \sum_{\mu}\ln H\left(-\frac{y^{\mu}\sum_i x_i^{\mu}\tanh \beta\theta_{i}} { \sqrt{\sum_{i}(1 - \tanh^2 \beta\theta_i) }}\right)$, 
	we can use the stochastic gradient descent (SGD)-based method to find a good trial distribution of weights, which may be a local or global minimum since the loss is a non-convex function.
	The gradients can be derived below, 
	\begin{equation}
	\begin{aligned}
		\frac{\partial \mathcal{L}}{\partial \theta_j^t} &= \beta(\sigma_j^t)^2\left(y \frac{x_j\sum_i(\sigma_i^t)^2+ \tanh(\beta\theta_j^t)\sum_ix_i\tanh(\beta\theta_i^t)}{(\sum_i(\sigma_i^t)^2 )^{\frac{3}{2}}}\right.\\
		&\left. \times H'\Biggl(-\frac{y\sum_ix_i\tanh(\beta\theta_i^t)}{\sqrt{\sum_i(1-\tanh^2(\beta\theta_i^t))}}\Biggr)\times H^{-1}\Biggl(-\frac{y\sum_ix_i\tanh(\beta\theta_i^t)}{\sqrt{\sum_i(1-\tanh^2(\beta\theta_i^t))}}\Biggr)\right)
	\end{aligned}
	\end{equation} 
	where $\sigma_j^2 = 1 - \tanh^2(\beta \theta_j)$ captures the weight uncertainty, the data index $\mu$ is neglected,
	and $t$ denotes the iterative time step. 
	The synaptic plasticity is thus modulated by the weight uncertainty,  which is biologically plausible~\cite{Alex-2021} and bears the similarity with other heuristic strategies~\cite{meta-2021,VCL-2020}.
	Another salient feature is that, provided that
	the uncertainty of a weight is small, this weight can be less plastic because of encoding important information of previous tasks.
	In addition, 
	the weight's synaptic plasticity is also tuned by the total uncertainty of the network, 
	$\sum_i(\sigma_i^t)^2 $, which plays a role of global regularization.
	
During the second-task learning, the posterior distribution of weights becomes
	\begin{equation}
		P(\bm{\xi}|\mathcal{\bm{D}}_1, \mathcal{\bm{D}}_2) = \frac{P(\mathcal{\bm{D}}_2|\bm{\xi}, \mathcal{\bm{D}}_1) P(\bm{\xi} | \mathcal{\bm{D}}_1 )}{P(\mathcal{\bm{D}}_2 | \mathcal{\bm{D}}_1)}.
	\end{equation}
	We assume that when the student learns the second task, the knowledge from the first task becomes a prior constraining the subsequent learning,
	i.e., $P(\bm{\xi} | \mathcal{D}_1 ) \simeq q_{\bm{\theta}^1}(\bm{\xi} ) $, 
	where $q_{\bm{\theta}^1}(\bm{\xi}  ) $ is the variational distribution after learning the first task. 
	We model the posterior of weights during learning of the second task as $
	q_{\bm{\theta}^2}(\bm{\xi}) = \prod_{i} \frac{e^{\beta\theta_{i}^2\xi_{i}}}{\cosh(\beta\theta_i^2)}$. Optimal 
	variational parameters
	can be obtained by minimizing the Kullback-Leibler (KL) divergence between variational distribution 
	and posterior distribution~\cite{Huang-2020},
	\begin{equation}\label{bound2}
		\begin{aligned} \bm{\theta} ^ { 2* } & = \arg \min _ { \bm{\theta}^2 } \mathbb{E}_{q_{\bm{\theta}^2}}\ln\frac{q_{\bm{\theta}^2}(\bm{\xi})} {P(\bm{\xi}| \mathcal{\bm{D}}_1, \mathcal{D}_2)} \\
		& = \arg \min _ { \bm{\theta}^2 } \mathbb{E}_{q_{\bm{\theta}^2}} \ln\frac{ q_{\bm{\theta}^2}(\bm{\xi})}{ q_{\bm{\theta}^1}(\bm{\xi})} - \mathbb{E}_{q_{\bm{\theta}^2}} \ln P(\mathcal{D}_2|\bm{\xi}) \\
		& \simeq  \arg \min _ { \bm{\theta}^2 } \mathrm{KL}\Bigl(q_{\bm{\theta}^2}(\bm{\xi})||q_{\bm{\theta}^1}(\bm{\xi})\Bigr) - \ln \mathbb{E}_{q_{\bm{\theta}^2}}  P(\mathcal{D}_2|\bm{\xi}),
		\end{aligned}
	\end{equation}
	where we discard $P(\mathcal{\bm{D}}_2 | \mathcal{\bm{D}}_1)$ because this term does not depend on the model parameters, and we use $P(\mathcal{\bm{D}}_2|\bm{\xi},\mathcal{\bm{D}}_1)=P(\mathcal{\bm{D}}_2|\bm{\xi})$, 
and we also approximate the objective function $\mathcal{L}$ by minimizing the lower bound of
	the KL divergence (in other words, we train the network to make the bound as tight as possible). 
	We remark here that one can also minimize the KL divergence between a trial probability $q_{\bm\theta}$ and the posterior
	$P(\bm{\xi}|\mathcal{\bm{D}}_1)$ for the first-task learning [see Eq.~\eqref{taskone}], which would require a prior probability of $\bm{\xi}$.
	Even if we set this prior to a uniform one, the system exhibits a similar learning behavior but the learning becomes harder as more data samples are required for reaching the same low 
	generalization error with the learning using Eq.~\eqref{taskone}. Therefore, we use Eq.~\eqref{taskone} as our first-task learning framework.
	The first term in Eq.~\eqref{bound2} is 
	a regularized term that makes the network to maintain the learned information of the first task. The second term
	is the expected log-likelihood term that leads the network to explain new data. The SGD-based method can then be
	applied to obtain the optimal solution (local or global minimum). The gradient can be computed as,
	\begin{equation}
	\begin{aligned}
		\frac{\partial \mathcal{L}}{\partial \theta_{j}^{2,t}}& = \beta(\sigma_{j}^{2,t})^2\left(\beta(\theta_{j}^{2,t}-\theta_{j}^{1})+y \frac{x_j\sum_i(\sigma_{i}^{2,t})^2+ \tanh(\beta\theta_{j}^{2,t})\sum_ix_i\tanh(\beta\theta_{i}^{2,t})}{(\sum_i(\sigma_{j}^{2,t})^2 )^{\frac{3}{2}}}\right.\\
		&\left.\times H'\Biggl(-\frac{y\sum_ix_i\tanh(\beta\theta_{i}^{2,t})}{\sqrt{\sum_i(1-\tanh^2(\beta\theta_{i}^{2,t}))}}\Biggr)\times H^{-1}\Biggl(-\frac{y\sum_ix_i\tanh(\beta\theta_{i}^{2,t})}{\sqrt{\sum_i(1-\tanh^2(\beta\theta_{i}^{2,t}))}}\Biggr)\right),
		\end{aligned}
	\end{equation}
	where $\mathcal{L}$ is the objective function to minimize in Eq.~\eqref{bound2}, and the data index is neglected and must refer to the task 2, and the gradient is still modulated by the weight uncertainty.
	Compared to the gradient of the first-task learning, the additional term comes from the KL
	divergence term. This term encourages the network to remember the first-task
	information. Therefore, this synaptic plasticity rule expresses the competition between old and new tasks (the second term).
	This trade-off allows the network to maintain the old
	knowledge but still adapt to the new task, thereby avoiding catastrophic forgetting to some extent.
\begin{figure}
\centering
     \includegraphics[bb=2 35 843 689,width=0.9\textwidth]{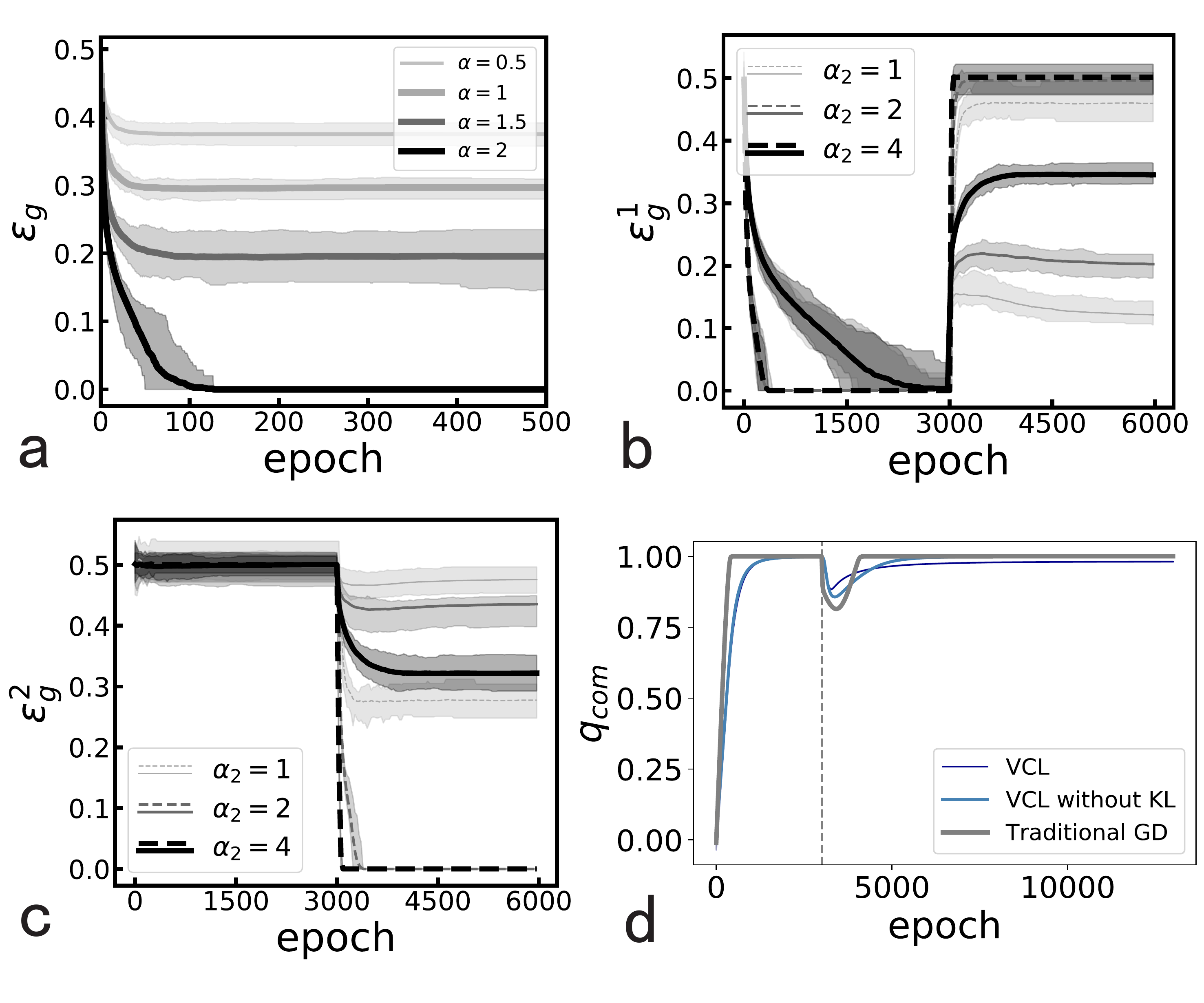}
  \caption{Learning performance of the toy model. (a) Test error of the first task with different $\alpha$. 
  (b) Test error of the first task. 
	(c) Test error of the second task. In (b,c), the task transition occurs at the $3000$-th epoch, and $r_0=0$.  $\alpha_1 = 2$, but $\alpha_2$ varies.
	Results are 
	averaged over 20 trials. (d) The overlap $q_{\rm com}$ between the student weights and the common part
	of both teachers (task similarity). $q_{\rm com}=\frac{1}{N_{\rm com}}\sum_{i}\hat{m}_i\hat{W}_i$, where
	$\hat{m}_i$ denotes the student's magnetization of the $i$-th synaptic weight and we choose $i$ such that $W_i^1=W_i^2$ ($=\hat{W}_i$).
	$N_{\rm com}$ denotes the total number of common weights in both teachers.
	In simulations, $r_0=0.5$, $\alpha_1 = 3$ and $\alpha_2=4$. For the traditional (full batch) GD algorithm, the optimization for two tasks is in the magnetization space (see Eq.~\eqref{task1obj} for the first task, and the second task has a similar form),
	and no KL divergence terms are used. The other two algorithms are implemented in the field ($\boldsymbol{\theta}$) space.
	The task switch (the dashed line) occurs at the 3000-th epoch. The network size $N=1\,000$. For (a,b,c), SGD is applied. For (d), the full batch GD is used. 
	The performance of the traditional SGD (dashed lines) is also shown in
	(b,c) for comparison.}
	\label{bperc}
\end{figure}

	We first show the simulation performance of our toy variational continual learning setting.
	 In Fig.~\ref{bperc} (a), with increasing amount of provided examples, the single-task learning performance improves.
	 In Fig.~\ref{bperc} (b), when the task 
	transition occurs, the test error of the first task increases, yet finally achieving a stable value across training.
	The test error of the second task 
	decreases, but can not achieve the error level that can be reached when the task is trained in isolation [Fig.~\ref{bperc}(c)].
	This is because the network does not forget the distinct characteristics of the first task completely, due to the regularization term.
	The lower test error would be achieved given more training examples for the second task.
	Figure~\ref{bperc} (d) illustrates the effect of the KL term, where we plot the overlap between the student inference and
	the common part of both teachers. Without the KL term, the overlap falls more sharply and then increases more rapidly, 
	while the presence of the KL term makes 
	the overlap change relatively slowly. This suggests that, the KL term makes the network tend 
	to protect the first task from a fast forgetting (see the poor performance of the traditional SGD in continual learning in Fig.~\ref{bperc}(b,c)]. The lower overlap (but still closer to one) allows more flexibility to balance
	the continual learning. 

We next derive the mean-field theory to evaluate analytically the continual learning performance.

\subsection{Mean-field theory: Franz-Parisi Potential}

Mean-field theory is a powerful tool for analyzing complex systems in statistical physics. In the previous section, 
we describe the variational method in training the binary perceptron to realize continual learning. In this section, we derive
 mean-field theory to analyze the variational continual learning. 
Instead of the local fields $\bm{\theta}^{1}$ and $\bm{\theta}^2$ (useful for practical training due to their unbounded values), 
we parameterize the variational distribution with weight-magnetization $m_{1,i} = \tanh{\beta\theta_{i}^1}$ and $m_{2,i} = \tanh{\beta\theta_{i}^2}$,
for the sake of analytical studies.
The variational distributions are specified respectively by
\begin{equation}
	\begin{aligned}
	Q_{\bm{m}_1}(\bm{\xi}) &= \prod_{i=1}^{N} \frac{1 + \xi_i m_{1,i}}{2},\\
	Q_{\bm{m}_2}(\bm{\xi}) &= \prod_{i=1}^{N} \frac{1 + \xi_i m_{2,i}}{2},
	\end{aligned}
\end{equation}
where $m_{1,i}$, $m_{2,i}$ $\in [-1,1]$ are the magnetization of the $i^{th}$ synaptic weight in the first and second task learning respectively. 
We perform the statistical mechanics analysis on the two-task learning, with the goal of extracting the role of model parameters (e.g., sample complexity, 
task similarity and so on) in the continual learning.

\subsubsection{The first-task analysis}

To perform the mean field theory analysis of the first-task learning, we define the loss function in the variational method as the Hamiltonian,
\begin{equation}\label{task1obj}
\mathcal { L } _ { 1 } ( \bm{m} ) = -\sum _ { \mu = 1 } ^ { M_1 } \ln H \left( - \frac { y^{\mu} \sum _ { i } m _ { i } x _ { i } ^ { 1 , \mu } } { \sqrt { \sum _ { i } \left( 1 - m _ { i } ^ { 2 } \right) } } \right),
\end{equation}
where $y^{\mu} = \operatorname { sign } \left( \sum _ { i } W _ { i } ^ { 1 } x _ { i } ^ { 1 , \mu } \right)$ is the label generated 
by the teacher network. The Boltzmann distribution reads
\begin{equation}
	P(\bm{m}) = \frac{1}{Z} e^{-\beta \mathcal{L}_1(\bm{m})},
\end{equation} 
where $\beta$ is an inverse temperature, $Z = \int_{\Omega} \prod_{i}dm_{i} e^{-\beta \mathcal{L}_1(\bm{m})}$ is the partition
function and the integral domain $\Omega= [-1,1]^{N}$. To obtain the equilibrium properties, we should first compute the disorder-averaged free energy (or the log-partition-function),
which can be achieved by using the replica trick.
The replica trick proceeds as $\langle\ln Z\rangle  =  \lim_{n\rightarrow 0} \frac{\ln\langle Z^n\rangle }{n}$, where $\langle\cdot\rangle$ denotes the average over the
quenched disorder. Then, we have
\begin{equation}
	\begin{split}
	\langle Z^{n} \rangle=\int_{\Omega^n} \prod_{a=1}^{n}   \prod_{i=1}^{N} \mathrm{d} m_{i}^{a} \left\langle\prod_{a=1}^{n}\prod_{\mu=1}^{M_1}  H^{\beta}\left(-\frac{\operatorname{sign}(\sum_{i} W_i^{1}x_i^{1,\mu})\sum_{i} m_{i}^a x_i^{1,\mu}}{\sqrt{\sum_{i}1-(m_{i}^a)^{2}}}\right)\right\rangle.
	\end{split}
\end{equation}
Under the replica symmetric (RS) Ans\"atz (detailed in the appendix~\ref{replica}), the free energy density at a given data density $\alpha = \frac{M}{N}$ is given by
\begin{equation}
	-\beta f_{\mathrm{RS}} = \lim_{n \rightarrow 0, N\rightarrow \infty} \frac{\ln\langle Z^n\rangle }{nN} = \lim_{n \rightarrow 0} 
	-\frac{1}{2}\left( \hat q_d q_d + (n-1) \hat q_{0} q_{0} \right) - \hat r_{1} r_{1} + \frac{\ln G_{\mathrm{S}}}{n} + \alpha_1\frac{\ln G_{\mathrm{E}}}{n},
\end{equation}
where
\begin{equation}
	\begin{split}
	G_\mathrm{E}&= \int \mathcal{D}z~2H\left(-\frac{r_{1}}{\sqrt{q_0-r_{1}^2}}z\right)\left( \int\mathcal{D}\sigma~H^{\beta}\left(-\frac{\sqrt{q_d-q_0} \sigma + \sqrt{q_0}z}{\sqrt{1-q_d}}\right)\right)^n,\\
	G_\mathrm{S} &= \int\mathcal{D}z \left(\int_{-1}^{+1}  \mathrm{d} m ~e^{\frac{1}{2} \hat{q}_{d} m^2 -  \frac{1}{2}\hat q_0 m^2 +\sqrt{\hat q_0} m z+\hat r m} \right)^n.
	\end{split}
\end{equation}

Order parameters are introduced as $q_0 = \frac{1}{N}\sum_{i} m_i^a m_i^b$ for $a\neq b$ indicating the overlap 
of different equilibrium states,  $q_d = \frac{1}{N}\sum_{i} m_i^a m_i^a$ indicating the self-overlap of states (the self-overlap 
relates to the size of valid weight space $\hat{\sigma} = 1-q_{d}$), and finally $r_1 = \frac{1}{N}\sum_{i}m_{i}W_i^1 $ indicating 
the overlap between the student's inference and the teacher's ground truth.
In practice, $q_d$ can be also estimated from the gradient descent dynamics during training, and we use $q_*$ to denote this measure, i.e., $q_*(t)=\frac{1}{N}\sum_i(m_i(t))^2$, and $\Delta m_i(t)\propto\partial_{m_i}\mathcal{L}_1(\bm{m})$.
$\{\hat{q}_0,\hat{q}_d,\hat{r}_1 \}$ are the conjugated order parameters introduced by the Fourier transform. 
These order parameters can be obtained by solving saddle point equations (detailed in appendix~\ref{replica}). To evaluate the learning performance, 
we define the generalization error $\epsilon_g^1=\langle\mathbb{E}_{\bm{x}^*}\Theta(-y^*\hat{y}^*)\rangle$, where $(\bm{x}^*,y^*)$ is the fresh data sample, and $\hat{y}^*$ is the student's prediction,
and $\langle\cdot\rangle$ denotes the disorder average.
The test error can be calculated as (see details in appendix~\ref{replica})
\begin{equation}\label{test error E}
	\epsilon_{g}^{1}= \int \mathcal{D}z ~ 2H\left(-\frac{p_{1}}{\sqrt{1-p_{1}^2}}z\right)\Theta(-z)=  \frac{1}{\pi}\arccos (p_1),
\end{equation}
where $p_1 = \frac{1}{N}\sum_{i}\text{sign}(m_i)W_i^1$ denotes the overlap between the decoded weights and the teacher's weights, and
can be obtained by solving the saddle point equations as well. 

\begin{figure}
\centering
     \includegraphics[bb=14 11 892 721,width=1.0\textwidth]{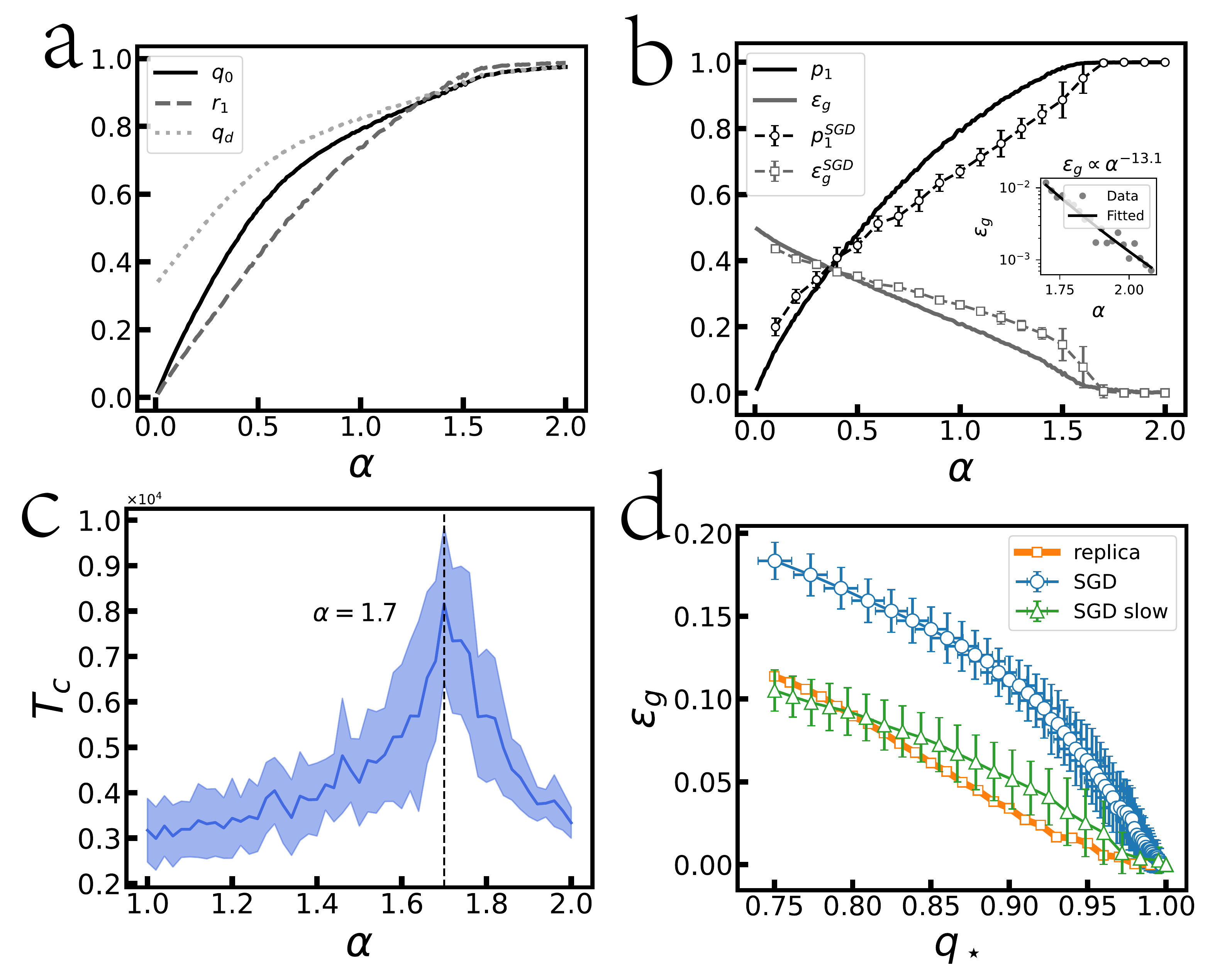}
  \caption{Mean-field results of the first-task learning compared with SGD simulations. 
		(a) The order parameters versus data density $\alpha$ ($\beta = 20$).
		(b) Generalization error versus data density ($\beta = 20$). The connected symbols
		 represent the result of SGD. 
		 The inset shows a power-law scaling for replica results when $\alpha\geq 1.7$.
		The symbol (Data) in the inset is the replica theory.
		(c) Convergence time step $T_c$ of the full batch gradient descent simulation. 
		$T_c$ is recorded when the drop of $\epsilon_g$ during $[T_c, T_c+1000]$ starts to be less than $0.0005$.
			The convergence time peaks at $\alpha=1.7$. The results are
			averaged over $20$ trials (network size $N = 3~000$).
		(d) Generalization error versus quenched $q_*$ ($\alpha=2$). The slow SGD means a rescale of
		the norm of $m$ to $q_\star$ after each update. 
		 The results of SGD are averaged over 20 trials (network size $N=1\,000$). }
	\label{rep01}
\end{figure}

We first show how order parameters change with respect to the data density $\alpha$. As defined, $q_d$ signals the size of the valid weight space. As the sample complexity increases,
the weight space shrinks down to a singe point representing the ground truth [Fig.~\ref{rep01} (a)]. As shown in Fig.~\ref{rep01} (b), the stochastic gradient descent dynamics results are approximately consistent with the theoretical predictions (at least qualitatively).
The deviations may be caused by the fact that the SGD could be trapped by local minima (suboptimal solutions) of the variational energy landscape.
But for a fixed $\alpha$, we can dynamically rescaled the norm of $\bm{m}$ after every update (slow type SGD~\cite{PRL-2018}), and compare the dynamical $\epsilon_g^1$ with its equilibrium counterpart (with the same value of $q_*$).
We find that the SGD results are comparable with the equilibrium predictions, at least qualitatively [Fig.~\ref{rep01} (d)]. The deviation may be caused by the finite size effects (see also the previous work~\cite{PRL-2018}).

In particular, Fig.~\ref{rep01} (b) reveals a continuous phase transition (from poor to perfect generalization)
in the variational parameter space (despite a binary perceptron learning considered in our setting).
Due to the numerical accuracy of the replica results at a large $\alpha$, we find that after $\alpha=1.7$, a power law scaling of the generalization error with
a large exponent [$\sim13.1$, see the inset of Fig.~\ref{rep01} (b)] is observed.
We remark that the SGD could reach zero error (perfect generalization) after $\alpha_c\simeq1.7$, while the replica result obtained at a large $\beta=20$ has a fast decay (lower than $10^{-2}$) after $\alpha_c$
[see the inset of Fig.~\ref{rep01} (b)].
It is expected that the replica prediction of the generalization error
will reach lower values with increasing $\beta$, which requires a huge number of Monte-Carlo samples to get an accurate estimate of the integral in Eq.~\eqref{test error E} and also in solving
the saddle-point equations (see details in appendix~\ref{replica}). The power-law fitting for the error below $10^{-3}$ may thus be unreliable. We conclude that in the zero temperature limit, the perfect generalization
is conjectured to be achievable, although a high $\beta$ may lead to replica symmetry breaking~\cite{PRL-2018}.
We could alternatively estimate the transition 
threshold by analyzing the convergence time of the learning algorithm [Fig.~\ref{rep01} (c)].
The convergence time is peaked at $\alpha_c\simeq1.7$, which is in stark contrast to the case of 
training in the direct binary-weight space~\cite{Sompolinsky-1990,Gyo-1990}, which leads to a discontinuous transition at $\alpha_c=1.245$ (a spinodal point locates at $\alpha_{\rm sp}=1.492$). This suggests that the variational learning erases the metastable regime where 
the poor generalization persists until the spinodal point.
Thus the variational framework bears optimization benefits for learning in neural networks with discrete weights.

\subsubsection{The second-task analysis}
Similarly, we specify the second-task Hamiltonian as follows,
\begin{equation}
	\mathcal{L}_2(\boldsymbol{m})=-\sum_{\mu=1}^{M_2} \ln H\left(-\frac{\operatorname{sign}(\sum_{i} W_i^{2}x_i^{2,\mu})\sum_{i} m_{i} x_i^{2,\mu}}{\sqrt{\sum_{i}\left(1-m_{i}^{2}\right)}}\right) + \sum_{i=1}^N \mathrm{KL}(Q_{m_i}\|Q_{m_{1,i}}).
\end{equation}
As explained before, the first term is a reconstruction term that maximizes the log-likelihood of the second-task data,
and the second term prevents the network from forgetting the previously acquired knowledge. Due to the regularization term, we have to treat the equilibrium analysis differently from the single task learning.
It is natural to put the analysis within the Franz-Parisi potential framework~\cite{Franz-1995,Huang-2014}. More precisely, we define the potential for the second-task learning as
\begin{equation}
    \Phi = \frac{1}{\tilde{Z}} \int_{\tilde{\Omega}}\prod_{i=1}^N \mathrm{d} \tilde{m}_{i}~e^{-\tilde{\beta} \mathcal{L}_1(\tilde{\boldsymbol{m}})} \ln\int_{\Omega} \prod_{i=1}^N \mathrm{d} m_{i}~  e^{-\beta \mathcal{L}_2(\boldsymbol{m},\tilde{\boldsymbol{m}})}.
\end{equation}
Taking the replica symmetric (RS) Ans\"atz, the disorder averaged potential is related to the following action $\mathcal{S}$ (see details in appendix~\ref{replica}), 
\begin{equation}
\begin{aligned}
	\mathcal{S} &= \lim_{n\rightarrow0}\lim_{s\rightarrow0}-\frac{1}{2}\left(n \hat{\tilde q}_{d} \tilde q_{d} + s(s-1) \hat{\tilde q}_{0} q_{0} \right) -\frac{1}{2}\left(s \hat q_{d} q_{d} + n(n-1) \hat q_{0} q_{0} \right) - n\hat{\tilde{r}}_1 \tilde{r}_1 - s\hat r_{2} r_{2}\\
	&+ \ln \mathcal{G}_{\mathrm{S}}
	+ \alpha_1\ln \mathcal{G}_{\mathrm{E}}^1+\alpha_2\ln\mathcal{G}_{\mathrm{E}}^2,
	\end{aligned}
\end{equation}
where 
\begin{equation}
	\begin{aligned}
		\mathcal{G}_{\mathrm{E}}^1
		&= \left\langle \prod_{a=1}^{n} H^{\tilde\beta}\left(-\frac{\operatorname{sign}({\tilde{v}_{1}}) \tilde u^{a}}{\sqrt{ 1-\tilde{q}_{aa}}}\right)\right\rangle\\
		&= \int \mathcal{D}z~2H\left(-\frac{\tilde{r}_{1}}{\sqrt{\tilde{q}_{0}-\tilde{r}_{1}^2}}z\right)\left( \int\mathcal{D}\sigma~H^{\tilde{\beta}}\left(-\frac{\sqrt{\tilde{q}_d-\tilde{q}_{0}} \sigma + \sqrt{\tilde{q}_{0}}z}{\sqrt{1-\tilde{q}_d}}\right)\right)^n,\\
	\end{aligned}
\end{equation}
and
\begin{equation}
	\begin{aligned}
		\mathcal{G}_{\mathrm{E}}^2
		&= \left\langle \prod_{c=1}^{s} H^{\beta}\left(-\frac{\operatorname{sign}({v_{2}}) u^{c}}{\sqrt{ 1-q_{cc}}}\right)\right\rangle\\
		&= \int \mathcal{D}z~2H\left(-\frac{{r_{2}}}{\sqrt{{q}_{0}-{r_{2}}^2}}z\right)\left( \int\mathcal{D}\sigma~H^{\beta}\left(-\frac{\sqrt{{q}_d-{q}_{0}} \sigma + \sqrt{{q}_{0}}z}{\sqrt{1-{q}_d}}\right)\right)^s,
	\end{aligned}
\end{equation}
and
\begin{equation}
	\begin{split}
	\mathcal{G}_{\rm S}&= \frac{1+r_0}{2} \int \mathcal{D} z_1  {\left( \int_{-1}^{+1} \mathrm{d}\tilde{m}~ e^{\tilde{\mathcal{I}}(\tilde{m},z_1)}  \right)^{n-1}} {\int_{-1}^{+1} \mathrm{d}\tilde{m}~  e^{\tilde{\mathcal{I}}(\tilde{m},z_1)} \int \mathcal{D} z_2 \left(\int_{-1}^{+1} \mathrm{d}m ~e^{\mathcal{J}^{+}(m, \tilde{m}, z_2)} \right)^s}\\
		&\quad + \frac{1-r_0}{2} \int \mathcal{D} z_1 {\left( \int_{-1}^{+1} \mathrm{d}\tilde{m}~ e^{\tilde{\mathcal{I}}(\tilde{m},z_1)}  \right)^{n-1}} {\int_{-1}^{+1} \mathrm{d}\tilde{m}~  e^{\tilde{\mathcal{I}}(\tilde{m},z_1)}  \int \mathcal{D} z_2\left(\int_{-1}^{+1} \mathrm{d}m ~e^{\mathcal{J}^{-}(m, \tilde{m}, z_2)} \right)^s}.
	\end{split}
\end{equation}
Note that, $\tilde{v}_1$ and $v_2$ are related to quenched disorder (see appendix~\ref{replica}), $\{q_{0}, r_2,q_{d}\}$ are order parameters in parallel to the first-task learning, while $\{\tilde{q}_{0}, \tilde{r}_1,\tilde{q}_{d}\}$ can be obtained by
solving the single-task saddle point equations (inherited from the first-task analysis). The functions $\tilde{\mathcal{I}}$ and $\mathcal{J}^{\pm}$ are defined in appendix~\ref{replica}.
The test error of the second task can be derived in the form, 
\begin{equation}\label{test error E1}
	\epsilon_{g}^{2}= \int \mathcal{D}z ~ 2H\left(-\frac{p_{2}}{\sqrt{1-p_{2}^2}}z\right)\Theta(-z)=  \frac{1}{\pi}\arccos (p_2),
\end{equation}
 where $p_2  = \frac{1}{N}\sum_{i}\text{sign}(m)W_{i}^2$ denote the overlap between the decoded weights and the second-teacher weights. Similarly, 
 the test error of the first-task after learning both tasks is given by
 \begin{equation}\label{test error E2}
 	\epsilon_{g}^{1}= \int \mathcal{D}z ~ 2H\left(-\frac{p_{1}}{\sqrt{1-p_{1}^2}}z\right)\Theta(-z)=  \frac{1}{\pi}\arccos (p_1),
 \end{equation}
where $p_1  = \frac{1}{N}\sum_{i}\text{sign}(m_{i})W_{i}^1$.

\begin{figure}
	\centering
	\includegraphics[bb=1 29 849 778,width=0.9\textwidth]{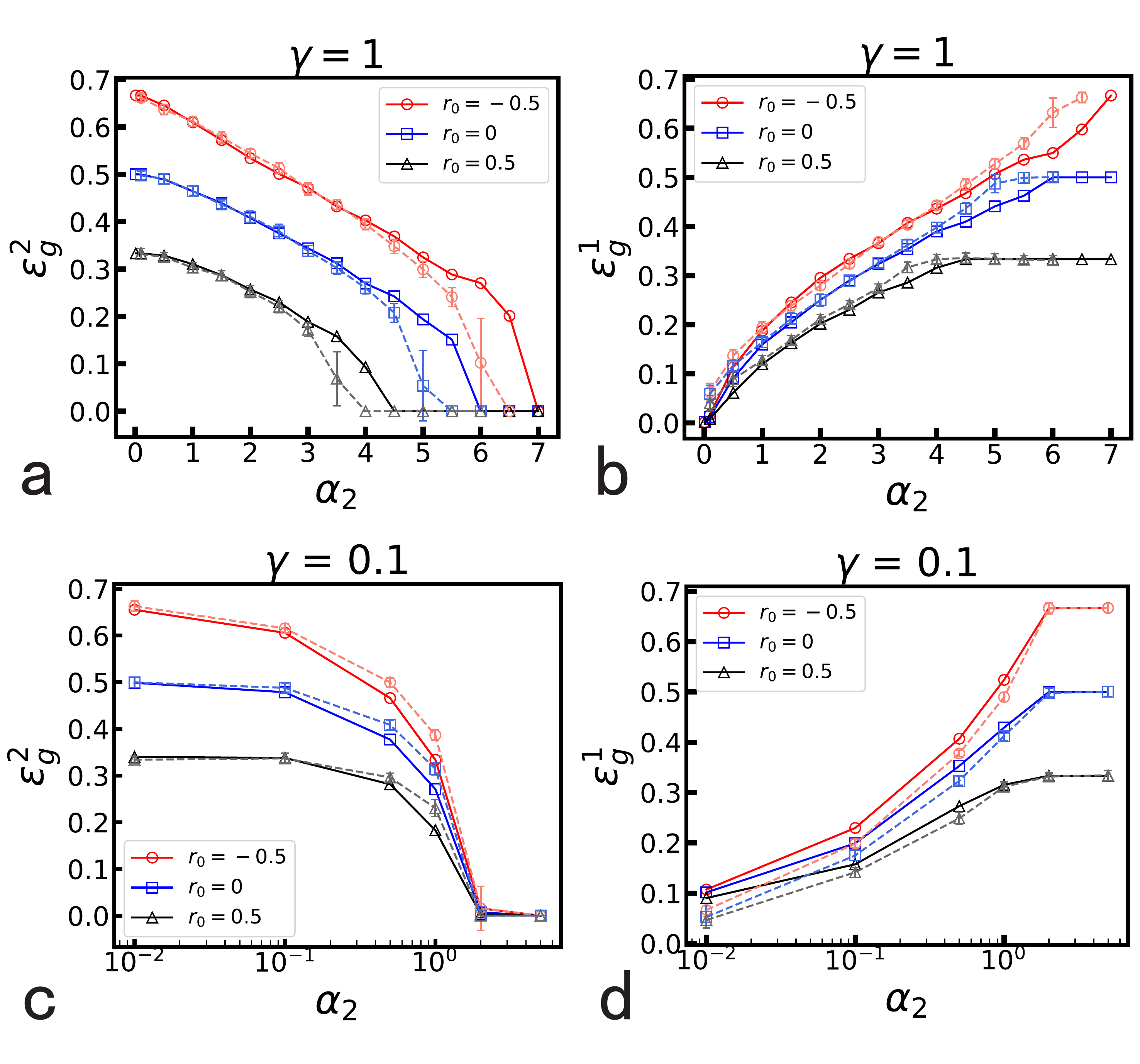}
	\caption{Generalization versus $\alpha$ for the variational continual learning. The symbols connected by dashed lines are
	simulation results of GD (twenty trials are averaged), while those connected by full lines are replica predictions.
	$\alpha_1=2$, $N=1\,000$, and different learning rates are used for different values of $\alpha_2$.
	Different task similarities are considered. The KL term is multiplied by a tuning factor $\gamma$.
	(a,b) $\gamma=1.0$. (c,d) $\gamma=0.1$.
	}\label{fig3}
\end{figure}

\begin{figure}
	\centering
	\includegraphics[bb=1 3 846 279,width=1.0\textwidth]{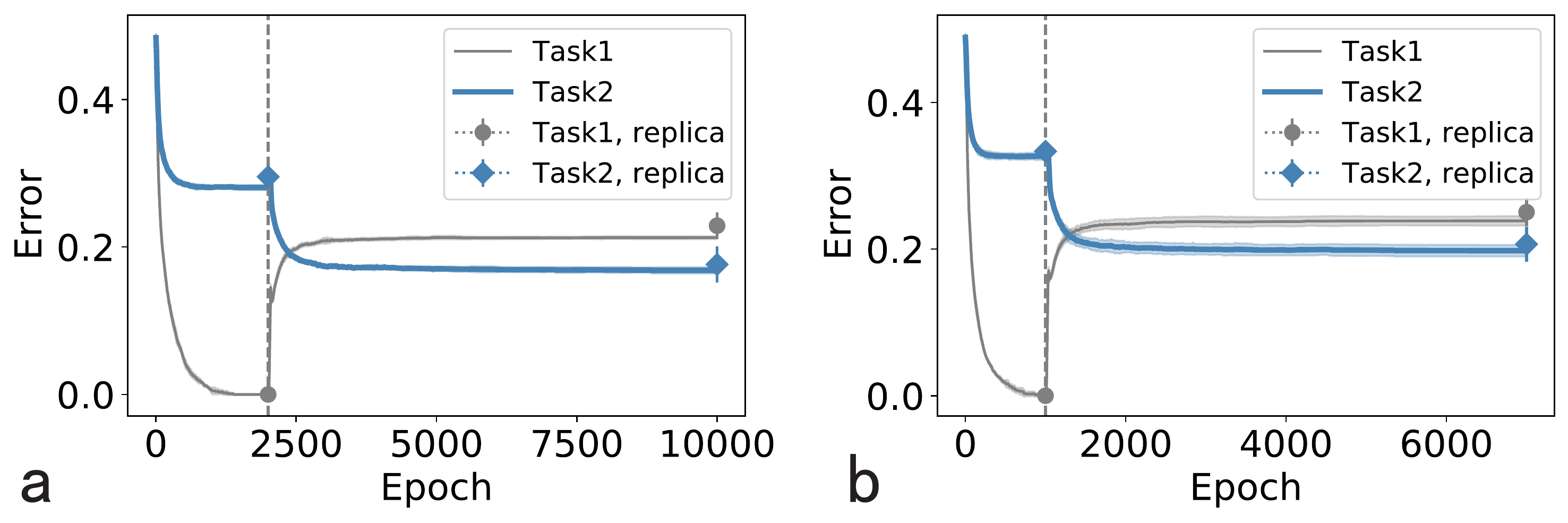}
	\caption{The comparison between replica results and simulation in perceptron. The simulation results are averaged over five independent trials. 
	The solid line shows the accuracy obtained from training the perceptron, while the symbols indicate
	the replica results. The dashed line indicates the task switch.	(a) $r_0=0.6$, and $\alpha_1 = \alpha_2=3.0$. (b) $r_0=0.5$, $\alpha_1 =4.0$, and $\alpha_2=3.0$.
	}\label{fig4}
\end{figure}

We finally study the theoretically predicted performances compared with numerical simulations. In Fig.~\ref{fig3} (a), 
we find that the task similarity strongly impacts the learning performance of the second task.  When $r_0$ takes a negative value, 
the learning becomes much harder, as more data examples are required to decrease the generalization error, while a positive
task similarity makes the learning of the second task easier. The SGD results match well with the theoretical prediction, except
for the region around the transition, which may call for longer simulation time in searching for good solutions. As expected, 
the generalization of the first task will increase during learning the second task [Fig. ~\ref{fig3} (b)], which is due to the 
fact that both tasks share a partial similarity (i.e., not completely the same). We also multiply the KL term by a factor $\gamma$, 
and study the effect of this term by tuning down this factor [e.g., $\gamma=0.1$ in Fig.~\ref{fig3} (c,d)]. We find that the learning
of the second task becomes fast as less data examples are required, and the critical value of $\alpha$ is also impacted.
Furthermore, the memorization of the first task is strongly degraded. This result is consisted with that found in Fig.~\ref{bperc} (d).

For a numerical verification of the mean-field replica theory, we train a perceptron
with the
number of synapses $N = 5000$. The learning rate equals to $0.001$ for the whole training process. 
If we use SGD, the size of a mini-batch is set to $32$. In the replica analysis, the hyperparameter $\beta_1=\beta_2=20$ for the first and second tasks.
Once the algorithm for the present task converges (e.g., the accuracy is stable), 
we shift the learning
to a new task. Figure~\ref{fig4} shows an excellent agreement between equilibrium predictions obtained by replica analysis and real training of perceptrons.

\section{Continual learning in deep neural networks}
Catastrophic forgetting is an unfavored property for deep neural networks applied to continual learning or multi-task learning.
 In this section, we extend the variational
methods for the toy binary perceptron to deep neural networks in classifying structured dataset.
\subsection{Variational learning principle}
Variational learning principle is a popular variational Bayesian framework applied in
a wide range of scenarios~\cite{VCL-2018,TaskAg-2018,Gal-2019,VCL-2020}, which focus mainly on deep networks with real-valued weights.
To learn a computationally efficient (binary weights) deep network, we adapt the variational principle to the continual learning, in theory and practical training, comparing the performance with that of 
the heuristic metaplasticity algorithm~\cite{meta-2021}, a unique available method for comparison in our current context.
Within this framework, the posterior of parameters $\mathbf{w}$ is
learned from $T$ continually presented datasets $\left\{\boldsymbol{x}_{t}^{(n)}, \boldsymbol{y}_{t}^{(n)}\right\}_{n=1}^{N_{t}}$, 
where $t$ denotes the task index ranging from $1$ to $T$, and $N_t$ denotes the size of dataset $t$. When the multi-task data examples are sequentially shown to the machine,
the posterior distribution of $\mathbf{w}$ is denoted as $p(\mathbf{w}| \mathcal{D}^k)$, 
after $k$-th training steps based on the dataset $\mathcal{D}^k$ (minibatch at the $k$-th step), can be calculated using the Bayes’ rule
as $p(\mathbf{w}|\mathcal{D}^k)=\frac{p(\mathcal{D}^k | \mathbf{w}) p(\mathbf{w})}{p(\mathcal{D}^k)}$. 
The prior $p(\mathbf{w})$ depends on the $(k-1)$-th step, which can be taken to be the posterior in
the previous training step $p(\mathbf{w}(k-1) |\mathcal{D}^{k-1})$ . 
Taken together, the posterior $p(\mathbf{w}\mid \mathcal{D})$ can be written as
\begin{equation}
p\left(\mathbf{w} | \mathcal{D}^k\right)=\frac{p\left(\mathcal{D}^k | \mathbf{w}\right) p\left(\mathbf{w}(k-1) | \mathcal{D}^{k-1}\right)}{p\left(\mathcal{D}^k\right)}.
\end{equation}
Unfortunately, the difficulty here is that the posterior is typically intractable for most of probabilistic models,
which thereby requires an application of the variational method. We approximate the true posterior with a tractable distribution 
parameterized by the variational parameter $\boldsymbol{\theta}$. By updating $\boldsymbol{\theta}$, we approach the target distribution
as close as possible.  
		
Given a simple trial probability distribution over the latent variable $\mathbf{w}$ parameterized 
by $\boldsymbol{\theta}$, i.e., $q_{\boldsymbol{\theta}}(\mathbf{w})$, the minimization of the
KL divergence between $q_{\boldsymbol{\theta}}(\mathbf{w})$ and $p(\mathbf{w}| \mathcal{D})$ results
in the following solution
\begin{equation}
\boldsymbol{\theta}^{*} =\arg \min _{\boldsymbol{\theta}} \mathrm{KL}\left[q_{\boldsymbol{\theta}}(\mathbf{w}) \| p(\mathbf{w}| \mathcal{D})\right].
\end{equation}
Therefore, we can define the loss function $\mathcal{L}$ as
\begin{equation}\label{essEq}
\begin{aligned}
	 \mathcal{L} &=\mathrm{KL}\left[q_{\boldsymbol{\theta}}(\mathbf{w}) \| p(\mathbf{w} | \mathcal{D})\right] = \mathcal{L}_{reg} + \mathcal{L}_{rec},
	\end{aligned}
\end{equation}
where we replace the prior $p(\mathbf{w})$ with the variational posterior in the previous 
time step $q_{\boldsymbol{\theta}^{k-1}}(\mathbf{w})$, and
we define the regularization term $\mathcal{L}_{reg} = \mathrm{KL}\left[q_{\boldsymbol{\theta}^k}(\mathbf{w}) \| q_{\boldsymbol{\theta}^{k-1}}(\mathbf{w})\right]$, 
and the reconstruction term $\mathcal{L}_{rec} = -\mathbb{E}_{q_{\boldsymbol{\theta}^k}(\mathbf{w})}\left[\ln p(\mathcal{D} | \mathbf{w})\right]$. 
The term $\mathcal{L}_{rec}$ computes the averaged log-likelihood of the network output, which can be crudely approximated
by considering a single sample $\mathbf{w}_s$ of $q_{\boldsymbol{\theta}^k}(\mathbf{w})$ from a rough Monte-Carlo sampling estimation. 
By computing gradients of $\boldsymbol{\theta}$ on this loss function $\mathcal{L}$, we arrive at
\begin{equation}
	\begin{aligned}
		\frac{\partial \mathcal{L}}{\partial \theta_{i}^k}  &= \sum_{w_{i}} \frac{\partial q_{\theta_i^k}(w_{i})}{\partial \theta_{i}^{k}}\left(1-\ln q_{\theta_i^k}(w_{i})-\ln q_{\theta_i^{k-1}}(w_{i})\right)\\
		&-\frac{\partial \ln p(\mathcal{D} \mid \mathbf{w}_s)}{\partial \theta_i^k}.
	\end{aligned}
\end{equation}
Learning of the variational parameter $\theta_i$ can be achieved by a gradient descent of
the objective function, i.e.,
\begin{equation}
	\theta_{i}^{k+1} = \theta_{i}^{k}-\eta \frac{\partial \mathcal{L}}{\partial \theta_i^k},
\end{equation}
where $\eta$ denotes the learning rate, and $\theta_i^k$ refers to the $i$-th connection in one layer a deep network (e.g., $\theta_{ij}^{l,k}$ for the connection $(ij)$
at layer $l$ below).

We consider a deep neural network with $L$ layers, and $N_l$ denotes the width of $l$th layer.
$w_{ij}^l$ indicates the weight connecting neuron $i$ at the upstream layer $l$ to neuron $j$ 
at the downstream layer $l+1$. The state of neuron $j$ at the $l+1$th layer $h_j^{l+1}$ is a
non-linear transformation of the preactivation $z_j^{l+1} = \frac{1}{\sqrt{N_l}}\sum_i w_{ij}^{l}h_{i}^{l}$. The 
transfer function $f(\cdot)$ for layers $l=1,2,\ldots,L-1$ is chosen to be the rectified linear 
unit (ReLU), which is defined as $f(z) = \max(0,z)$. For the output layer, the softmax 
function $h_{k}=e^{z_{k}} / \sum_{i} e^{z_{i}}$ is used, specifying the probability
over all classes of the input images, where $z_i$ is the preactivation of neuron $i$ at the output
layer. The supervised learning is considered, where $\hat{h}_k$ indicates the target of $h_k^L$, and
the cross entropy $\mathcal{L}_{\rm ce}=-\sum_i\hat{h}_i\ln h_i$ is used as a cost function corresponding to $\mathcal{L}_{rec}$.
In our setting, a double-peak distribution is applied to model the binary weight as 
$q_{ \theta_{i j}^l}\left(w_{i j}^l\right)=\frac{e^{\beta w_{i j}^{l} \theta_{i j}^{l}}}{e^{\beta \theta_{i j}^{l}}+e^{-\beta \theta_{i j}^{l}}}$, 
where $\beta$ is a hyperparameter, and the field-like parameter $\theta_{ij}$ controls the probability distribution of
$w_{ij}$ as $q_{\theta_{i j}^l}\left(+1\right)=\frac{e^{\beta\theta_{i j}^{l}}}{e^{\beta \theta_{i j}^{l}}+e^{-\beta \theta_{i j}^{l}}}$
and $q_{\theta_{i j}^l}\left(-1\right)=\frac{e^{-\beta\theta_{i j}^{l}}}{e^{\beta \theta_{i j}^{l}}+e^{-\beta \theta_{i j}^{l}}}$. 
Therefore, the gradients of $\boldsymbol{\theta}$ on $\mathcal{L}_{reg}$ can be computed as
\begin{equation}
	\begin{aligned}
	&\frac{\partial \mathrm{KL}\left[q_{{\theta}_{ij}^{l,k}}({w}_{ij}^{l,k}) \| q_{{\theta}_{ij}^{l,k-1}}({w}_{ij}^{l,k-1})\right]}{\partial \theta_{i j}^{l,k}}\\
	&=\beta^{2}\left(\theta_{i j}^{l,k}-\theta_{i j}^{l,k-1}\right)\left(\sigma_{i j}^{l,k}\right)^{2}=\beta^{2}\left(\sigma_{i j}^{l,k}\right)^{2} \Delta_{i j}^{l, k},
	\end{aligned}
\end{equation}
where the superscript $l$ and $k$ denote the layer index and iteration step respectively,
and we define $\Delta_{i j}^{l, k}$ as the increments of the variational parameter $\Delta_{i j}^{l,k}=\theta_{i j}^{l,k}-\theta_{i j}^{l, k-1}$
between two successive steps. $\left(\sigma_{i j}^{l,k}\right)^{2}$ indicates 
the variance of $w_{ij}^l$ as $\left(\sigma_{i j}^{l, k}\right)^{2}=1-\tanh ^{2}\left(\beta \theta_{i j}^{l, k}\right)$, 
and thus captures the synaptic uncertainty. 

To derive the gradients of $\mathcal{L}_{rec}$,  we apply the mean-field method~\cite{Li-2020}. 
The first and second moments of $w_{ij}^l$ are given by $\mu_{i j}^{l}=\langle w_{i j}^{l}\rangle=\tanh \left(\beta \theta_{i j}^{l}\right)$ 
and $\left(\sigma_{i j}^{l}\right)^{2} = 1-\left(\mu_{i j}^{l}\right)^{2}$, respectively. Given that
the width of layer is large, the central-limit theorem indicates that the preactivation
$z_{j}^{l+1}$ follows a Gaussian distribution $\mathcal{N}(z\mid m_{j}^{l+1};v_{j}^{l+1})$, where the mean and variance are given below,
\begin{equation}
	\begin{aligned}
		m_{j}^{l+1} &=\langle z_{i}^{l}\rangle=\frac{1}{\sqrt{N_{l}}} \sum_{j} \mu_{ij}^{l} h_{i}^{l} \\
		\left(v_{j}^{l+1}\right)^{2} &=\langle \left(z_{j}^{l+1}\right)^{2}\rangle-\langle z_{j}^{l+1}\rangle^{2}=\frac{1}{N_{l}} \sum_{j}\left(\sigma_{i j}^{l}\right)^{2}\left(h_{i}^{l}\right)^{2}.
	\end{aligned}
	\label{eq6}
\end{equation}
Therefore we write the preactivation as $z_{i}^{l}=m_{i}^{l}+\epsilon_{i}^{l} v_{i}^{l}$, where $\epsilon_i^l$ denotes 
a standard Gaussian variable relying on the layer and weight-component index. Then, we can
compute the gradients as follows,
\begin{equation}
	\frac{\partial \mathcal{L}_{rec}}{\partial \theta_{i j}^{l}}=\frac{\partial  \mathcal{L}_{rec}}{\partial z_{j}^{l+1}} \frac{\partial z_{j}^{l+1}}{\partial \theta_{i j}^{l}}=\mathcal{K}_{j}^{l+1}\left(\frac{\partial m_{j}^{l+1}}{\partial \theta_{i j}^{l}}+\epsilon_{j}^{l+1} \frac{\partial v_{j}^{l+1}}{\partial \theta_{i j}^{l}}\right),
\end{equation}
where we have defined $\mathcal{K}_{j}^{l+1} = \frac{\partial  \mathcal{L}_{rec}}{\partial z_{j}^{l+1}}$, 
which could be solved using the chain rule. The term $\frac{\partial m_{j}^{l+1}}{\partial \theta_{i j}^{l}}$ 
and $ \frac{\partial v_{j}^{l+1}}{\partial \theta_{i j}^{l}}$ can be directly derived from Eq.\eqref{eq6}, as shown below,
\begin{equation}
	\begin{aligned}
		\begin{gathered}
			\frac{\partial m_{j}^{l+1}}{\partial \theta_{i j}^{l}}=\frac{1}{\sqrt{N_{l}}} \beta h_{i}^{l}\left(\sigma_{i j}^{l}\right)^{2}, \\
		 \frac{\partial v_{j}^{l+1}}{\partial \theta_{i j}^{l}}=-\beta \frac{\left(h_{i}^{l}\right)^{2}}{N_{l} v_{j}^{l+1}} \mu_{i j}^{l}\left(\sigma_{i j}^{l}\right)^{2},
		\end{gathered}
	\end{aligned}
\end{equation}
and $\mathcal{K}_i^l$ can be estimated by the chain rule from the value at the top layer, i.e.,
\begin{equation}
 \mathcal{K}_{i}^{l}=\sum_{j} \mathcal{K}_j^{l+1}\left(\frac{1}{\sqrt{N_{l}}} \mu_{i j}^{l}+\frac{\epsilon_{j}^{l+1}}{N_{l} \sqrt{\left(v_{j}^{l+1}\right)^{2}}}\left(\sigma_{i j}^{l}\right)^{2} h_{i}^{l}\right) f^{\prime}\left(z_{i}^{l}\right),
\end{equation}
where $f^{\prime}(\cdot)$ is the derivative of transfer function, and on the top layer, $\mathcal{K}_i^L$ can be directly
estimated as $\mathcal{K}_i^L = -\hat{h}_i(1-{h}_i^L)$. Taken together, the total gradients on the loss 
function $\mathcal{L}$ take the form as
\begin{equation}\label{vcldl}
	\frac{\partial \mathcal{L}}{\partial \theta_{i j}^{l,k}}=\beta\left(\sigma_{i j}^{l,k}\right)^{2}\left(\beta \Delta_{i j}^{l,k} + \delta_{ij}^{l,k}\right), 
\end{equation}
where we add the step index $k$ as $\delta_{ij}^{l,k}=\mathcal{K}_j^{l+1}\left(\frac{1}{\sqrt{N_{l}}} h_{i}^{l}-\frac{\epsilon_{j}^{l+1}}{N_{l} v_{j}^{l+1}}\left(h_{i}^{l}\right)^{2} \mu_{i j}^{l}\right)$.
It can be clearly seen that the variance of $w_{ij}^l$ at the iteration step $k$ together with 
the inverse temperature $\beta\left(\sigma_{i j}^{l,k}\right)^{2}$ tunes the learning rate $\eta$,
where a larger variance leads to larger gradients in the iteration step $k$. In addition, the regularization term
$\beta \Delta_{i j}^{l,k-1}$ measures the similarity between the variational posterior probabilities across successive learning
steps, which regulates the distance from current guess to previous one, providing a principled way to use 
the information from previous task knowledges. Therefore, this variational continual learning can be used in scenarios where task 
boundaries are not available~\cite{TaskAg-2018}, which is also more cognitively plausible from our humans' learning experiences. 
Hereafter, we call this variational continual learning scheme as VCL.

We emphasize the relationship between the VCL used in toy model analysis in previous section and that used for practical continual deep learning in this section.
In essence, the VCL in these two sections bears the same principle [see Eq.~\eqref{essEq}].
In the toy model analysis, we specify the task boundary, which allows us to derive the Franz-Parisi potential of the continual learning. However, in a practical training,
a task agnostic training is favored (like humans), which is exactly captured in Eq.~\eqref{vcldl}. Therefore, the last training step acts as a reference in the language of 
the Franz-Parisi framework, i.e., the learning of the next step can be described by an equilibrium system with an anchored external preference.
Furthermore, in the toy model analysis, we set the hyperparameter $\beta$ for both tasks the same value. In the practical deep learning, $\beta$ is allowed to increase
with epoch [one example is shown in Fig.~\ref{fig6} (a)]. 

In particular, our learning protocol emphasizes how synaptic uncertainty tunes the continual learning, and thus provides a principled way to understand engineering heuristics~\cite{CLAW-2020,VCL-2020}
and neuroscience inspired heuristics~\cite{PNAS-2017,PNAS-2018,meta-2021}. For the deep networks with binary weights, the previous work 
uses discretization operation of a continuous weight, surrogate gradient and a metaplasticity function (see details in Appendix~\ref{exp}),
while our VCL does not require these
tricks. In addition, other heuristic strategies such as elastic weight consolidation and its variants~\cite{PNAS-2017} can be also unified in our current framework.
For example, the part $\mathcal{L}_{reg}$ can
be approximated by a term involving the Fisher information matrix as
\begin{equation}
 F(\boldsymbol{\theta})=\mathbb{E}_{q_{\boldsymbol{\theta}}(\mathbf{w})}\left(\left(\frac{\partial \ln q_{\boldsymbol{\theta}}(\mathbf{w})}{\partial \boldsymbol{\theta}}\right)\left(\frac{\partial \ln q_{\boldsymbol{\theta}}(\mathbf{w})}{\partial \boldsymbol{\theta}}\right)^{\top}\right).
\end{equation}
and then
\begin{equation}
	\mathcal{L}_{reg}\approx \frac{1}{2}\left(\boldsymbol{\theta}^{k}-\boldsymbol{\theta}^{k-1}\right)^{\top} F\left(\boldsymbol{\theta}^{k-1}\right)\left(\boldsymbol{\theta}^{k}-\boldsymbol{\theta}^{k-1}\right),
\end{equation}
where we take only the diagonal elements of the Fisher information matrix $F\left(\boldsymbol{\theta}^{k-1}\right) \approx \beta^2 (\boldsymbol{\sigma}^{k-1})^2$, recovering the elastic weight consolidation algorithm.
The technical proof is given in appendix~\ref{proof}.

\begin{figure}
	\centering
	\includegraphics[bb=1 2 842 268,width=1.0\textwidth]{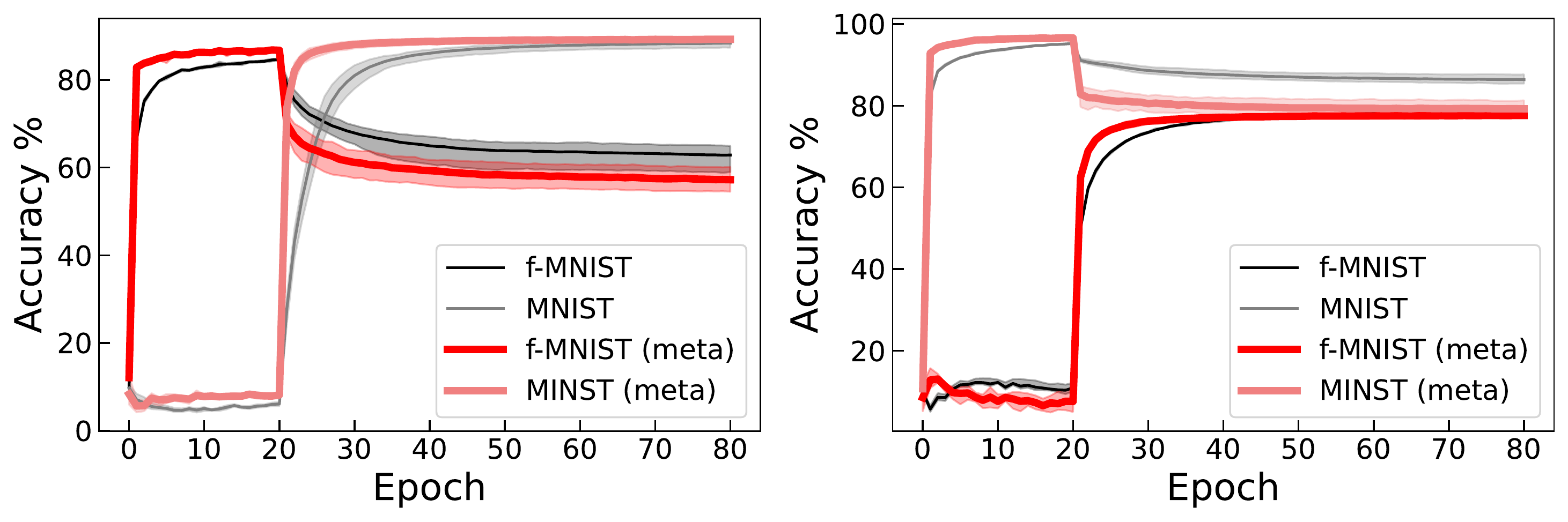}
	\caption{Continual learning on MNIST and Fashion-MNIST (f-MNIST) datasets.
	The network has the architecture $[784,400,200,10]$, each number 
	indicates the layer width. The results are averaged over five independent trials. (Left panel) The training order is f-MNIST first and then
	MNIST. $\beta_1=1$ for the first task and $\beta_2 = 12.5$ for the second one
	in the VCL setting. $m_1 = 0.5$ for the first task and $m_2 = 0.9$ for the 
	second in the metaplasticity algorithm. (Right panel) MNIST is applied first followed by
	the f-MNIST dataset. The same network architecture is used as (a), while $\beta_1=1$, 
	$\beta_2 = 17.5$,  $m_1 = 0.5$ and $m_2 = 0.7$.
	}\label{fig5}
\end{figure}

\begin{figure}
	\centering
	\includegraphics[bb=1 2 842 268,width=1.0\textwidth]{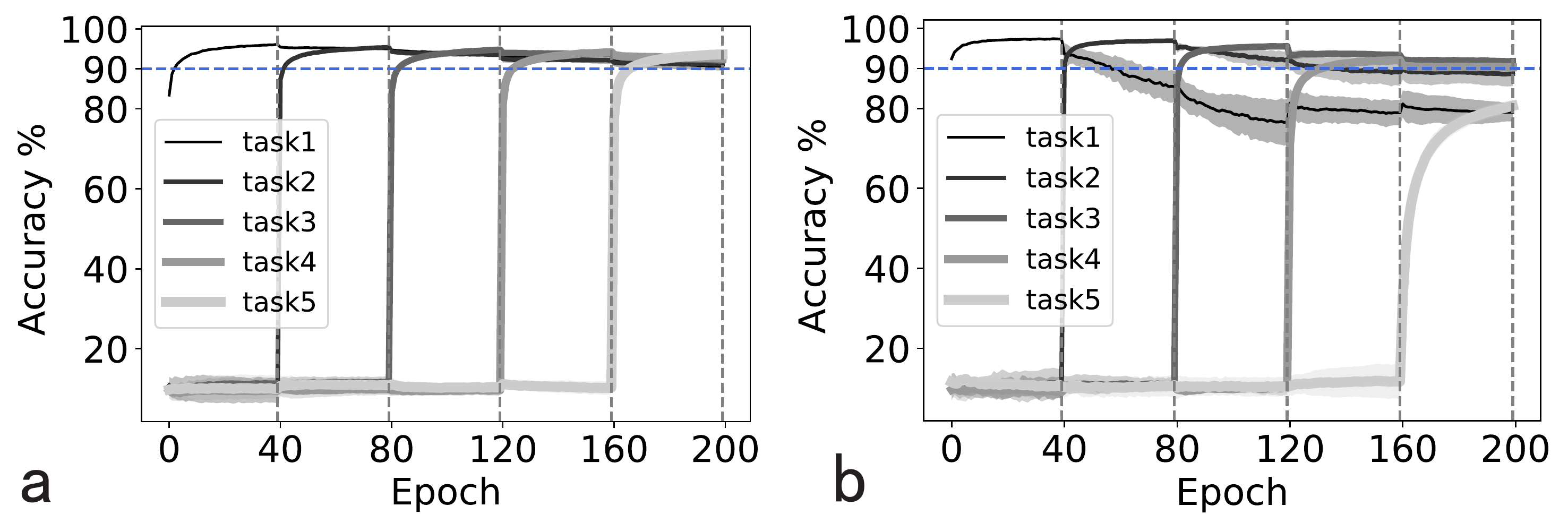}
	\caption{Continual learning of permuted MNIST learning tasks. Five sequential tasks are 
	considered, and each is trained for 40 epochs. The network has the architecture $[784,512,512,10]$, each number 
	indicates the layer width. The results are averaged over five independent trials. (a) Test accuracy based on
	VCL, $\beta_\ell=a\tanh(\delta+b\ell/M)$, where $\ell$ denotes the epoch index, and $M$ is the total number of epochs. We use $a=10.0$, $\delta=0.1$, and $b=2.0$. 
	(b) Test accuracy of the metaplasticity algorithm, for which $m=0.43$ for all the tasks.
	}\label{fig6}
\end{figure}

\begin{figure}
	\centering
	\includegraphics[bb=1 2 506 269,scale=0.7]{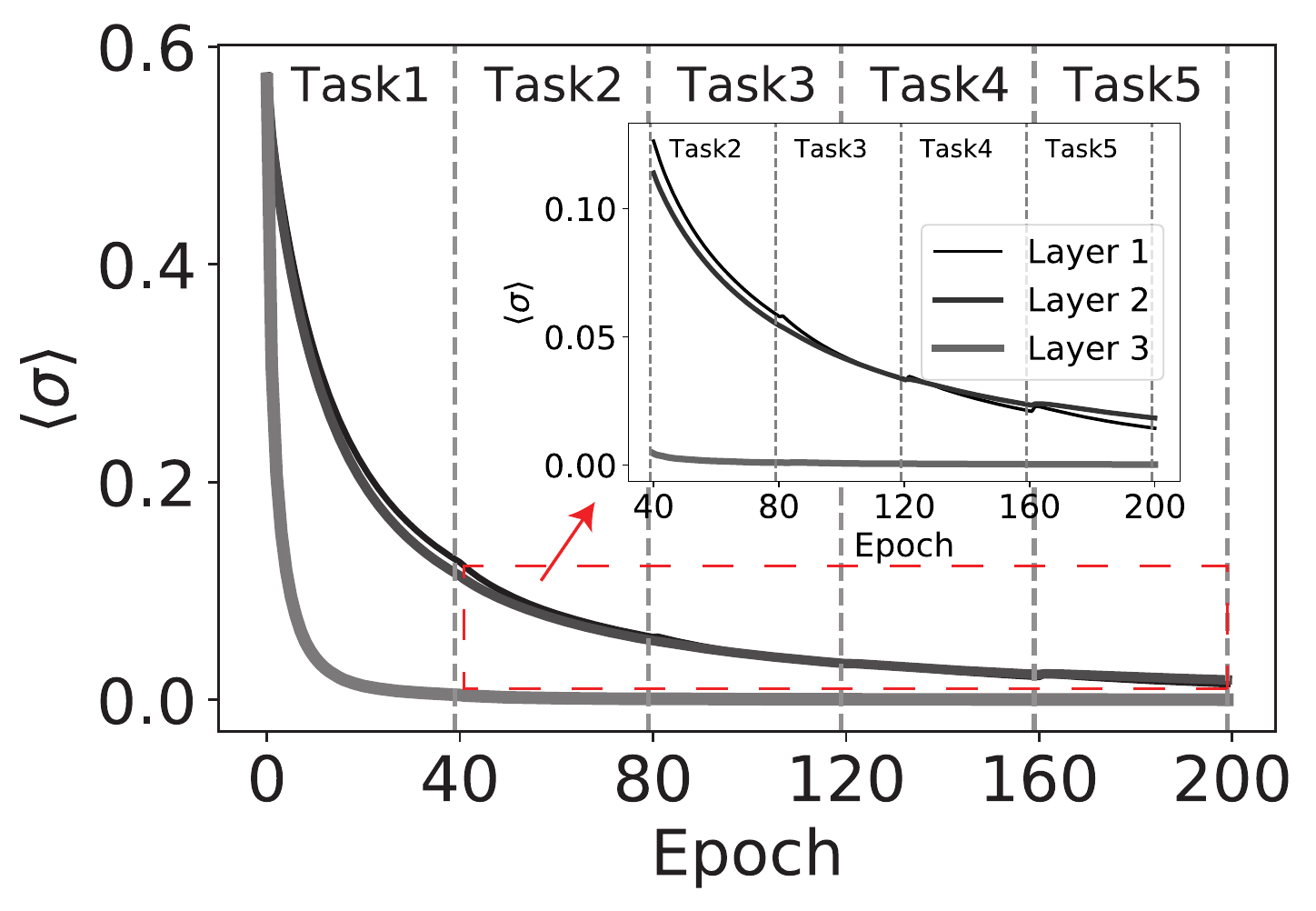}
	\caption{The averaged level of synaptic uncertainty ($\langle{\sigma}\rangle$) evolves
	through training for all the layers, and the inset shows the details
	of the training stage after $40$ epochs. The network architecture is the same
	as that of Fig.~\ref{fig6}.
	}\label{fig7}
\end{figure}

\subsection{Learning performance and roles of synaptic uncertainty}
In this section, we compare the performance of VCL and the metaplasticity algorithm for neural networks with binary weights.
Algorithmic details are given in appendix~\ref{exp}. We consider two tasks first--- sequential learning of MNIST and Fashion-MNIST (f-MNIST) datasets~\cite{meta-2021}, and
this setting requires the network
to sequentially learn from two datasets: MNIST and f-MNIST. We next consider a popular continual learning benchmark, namely
the permuted MNIST learning
task~\cite{PNAS-2017}. The permuted MNIST learning task
is composed of continual learning of several datasets, and each task contains labeled images of a fixed random spatial permutation of pixels.

Figure~\ref{fig5} shows that for both training orders (f-MNIST first or MNIST first), 
VCL achieves a much better performance than that of the metaplasticity algorithm (a shorthand as meta), 
showing the benefit of less forgetting of learned tasks and thus better performance for new coming 
tasks. The same phenomenon can be also observed in Fig.~\ref{fig6}, where five permuted MNIST datasets are sequentially presented to
the network. Networks trained with VCL [Fig.~\ref{fig6} (a)] are learning better and
forgetting less previous task knowledge compared with those trained with meta [Fig.~\ref{fig6} (b)]. The plot shows the classification accuracy for task $t$ after learning tasks $t'\geq t$.
A perfect continual learning must provide high accuracy for task $t$, and moreover preserve the performance even when subsequent tasks are learned, which is 
independent of the training time of each task (e.g., increasing the training time to $100$ epochs).

We also plot the evolution of synaptic uncertainty by calculating the average $\langle\sigma\rangle=\frac{1}{\#{\rm weights}}\sum_{(ij)}\sigma_{ij}^2$ for every 
epoch and every layer. As expected, the mean uncertainty decreases during the continual learning. However, the level rises with a minor magnitude after a task switch, but then drops again. 
In addition, the reduction of the uncertainty is also evident for upstream layers, indicating that these layers tend to freeze most of weights, or make them less plastic.
In contrast, the last layer maintain a low level of synaptic uncertainty for reading out the key category information. 
We also observe that the synaptic plasticity with a larger uncertainty has a larger contribution to how strong the KL divergence should change, which thereby plays
an important role in minimizing the overall objective.
To conclude, the synaptic variance is a key quantity determining 
the behavior of continual learning. The VCL can adjust the synaptic resources during sequentially learning of multiple tasks.

\section{Conclusion}
In this study, we focus on the continual learning in deep (or shallow) neural networks with binary weights.
Recent works already argued that the variational training is effective in neural networks of real-valued weights~\cite{TaskAg-2018,VCL-2018,VCL-2020}, and a brain-inspired metaplasticity method
is also effective in training binary neural networks~\cite{meta-2021}. However, how to unify these diverse strategies within a statistical physics model is challenging. 
Here, we propose a variational mean-field framework to incorporate synaptic uncertainty, task-knowledge transfer and mean-field potential for multi-task learning.
First, we argue that the synaptic uncertainty plays a key role in modulating continual learning performance, through the lens of variational weight distribution.
Specifically, the synaptic variance becomes a modulating factor in the synaptic plasticity rules, based on our theory. Second, the task-knowledge transfer can be interpreted in
physics. The knowledge from the previous task behaves as a reference configuration in the Franz-Parisi potential formula~\cite{Franz-1995,Huang-2014}, an anchor for learning new knowledge.
The learning of new task can thus be described by an equilibrium system with an anchored external preference. The derived theory matches well the numerical simulations using
stochastic gradient descent algorithms.

Our theory of variational continual learning also predicts that a single-task learning exhibits a continuous transition with increasing amount of data (sample complexity),
which is in stark contrast to the previous findings in mean-field theory of generalization (in the direct discrete or continuous weight space)~\cite{Gyo-1990,Sompolinsky-1990}.
This new theoretical prediction suggests that the current variational continual learning proves efficient in practical learning, since a trapping by metastable states is absent.
We remark that this absence of a first-order transition holds only for shallow networks. It is thus interesting to extend our theoretical analysis to multi-layered networks to see if 
this conclusion is present or not.

We finally demonstrate that our framework can be applied to continual learning of real datasets, achieving similar or even better
performances with those obtained by heuristic strategies, such as metaplasticity. Therefore, this work can be a promising starting point to explore further the important yet challenging question of 
how to build theoretically-grounded neural representation that helps an intelligent agent avoid catastrophic forgetting and adapt continuously to new tasks, based on 
accumulated knowledges from previous tasks.

\appendix
\section{Algorithmic details}\label{exp}
In this section, we provide the details of metaplasticity algorithm and VCL, 
which are compared in the main text. The pseudocode of the metaplasticity algorithm~\cite{meta-2021} is summarized in Algorithm~\ref{meta}.
\begin{algorithm}[H]
	\caption{Metaplasticity continual learning}\label{meta}
	\begin{algorithmic}[1]  
		\State {Input: $h^{l-1}_{j}$, $w_{ij}^l = \operatorname{sign}(a_{ij}^l)$; meta parameter $m$; learning rate $\eta$}
		\State{Feedforward propagation: $z_{i}^{l}=\frac{1}{\sqrt{N^{l-1}}}\sum_{j} w_{j i}^{l} h_{j}^{l-1},
			h_{i}^{l}=f\left(z_{i}^{l}\right),  f(x)=\operatorname{ReLu}\left(x\right)$;}
		\State{Backpropagation: $\frac{\partial \mathcal{L}}{\partial w_{i j}^{l}}=\frac{\partial \mathcal{L}}{\partial z_{j}^{l}} \frac{\partial z_{j}^{l}}{\partial w_{i j}^{l}}=\left(\frac{1}{\sqrt{N^{l-1}}}\right)\sum_{k} \frac{\partial \mathcal{L}}{\partial z_{k}^{l+1}} w_{j k}^{l+1} f^{\prime}\left(z_{j}^{l}\right) h_{i}^{l-1}$;}
		\State{Parameter update: $\left(a_{i j}^{l}\right)^{k+1}-\left(a_{i j}^{l}\right)^{k}=-\eta \left(1-\zeta\tanh^2(m\left(a_{i j}^{l}\right)^{k})\right)\frac{\partial \mathcal{L}}{\partial w_{i j}^{l}}$, where $\zeta=\frac{1}{2}\Biggl[\operatorname{sign}\Biggl(w_{i j}^{l}\frac{\partial \mathcal{L}}{\partial w_{i j}^{l}}\Biggr)+1\Biggr]$.}
	\end{algorithmic}
\end{algorithm}
In the meta algorithm, $\mathcal{L}$ denotes the loss function, and $a_{ij}^l$ denotes the latent real-valued weights underlying binary counterpart, and the core idea is the introduction of 
a modulation function $f_{\rm meta}(m,x)=1-\tanh^2(mx)$ which is a decreasing function of $|x|$ (or the absolute value of the hidden weights).
This modulation called metaplasticity makes the hidden weight change less likely if the corresponding magnitude are growing (consolidation of some useful information, expressed as the $\zeta$ factor).
Therefore, this metaplasticity can be heuristically thought of as a sort of weight consolidation. 
In contrast, our VCL gives rise to an alternative modulation related to the synaptic uncertainty, thereby bearing a more solid theoretical ground.
In addition, the gradient with respect to a discrete weight value in the meta algorithm is ill-defined, which does not appear in our VCL.
We remark that this
algorithm is sensitive to the network size; if the size of network is not big enough, this algorithm may
fail to give satisfied learning performance.

Our VCL algorithm is summarized in the pseudocode~\ref{vcl}.
 In the main text, 
we use learning rate---$0.01$ and mini-batch size---$64$ for both tasks. 
Note that the mean and variance of the preactivation $\boldsymbol{z}$ is computed for each single data sample, 
	given the statistics of the weight. All codes to reproduce our results in the main text are available at the Github link~\cite{code}.
\begin{algorithm}[H]
		\caption{VCL algorithm}\label{vcl}
		\begin{algorithmic}[1]   
			\State {Input: single sample $\mathbf{x}\in \mathbb{R}^{N_0}$, $\mathbf{w}^{l}\in \mathbb{R}^{N_l\times N_{l+1}}$ from the distribution $q_{\theta_{i j}^l}\left(w_{i j}^l\right)=\frac{e^{\beta w_{i j}^l \theta_{i j}^l}}{e^{\beta \theta_{i j}^l}+e^{-\beta \theta_{i j}^l}}$;}
			\State{Compute the mean and variance of elements $\mathbf{w}^l: \boldsymbol{\mu}^{l},\left(\boldsymbol{\sigma}^{l}\right)^{2}$;}
			\State{Compute the mean $\boldsymbol{m}^{l}$ and variance $\left(\boldsymbol{v}^{l}\right)^{2}$ of preactivation $\boldsymbol{z}^l = \frac{1}{\sqrt{N_l}}(\mathbf{w}^{l-1})^\top \mathbf{h}^{l-1}$, $\mathbf{h}^0 = \mathbf{x}$;}
			\State{Sample $\boldsymbol{\epsilon}^l\in \mathbb{R}^{N_{l}}$ independently from $\mathcal{N} \sim(0,1)$;}
			\State{Output: $\boldsymbol{z}^{l}=(\boldsymbol{m}^{l}+\boldsymbol{\epsilon}^{l} \odot \boldsymbol{v}^{l}), \boldsymbol{h}^{l}=f\left(\boldsymbol{z}^{l}\right)$;}
			\State{Parameter update: $\theta_{ij}^{l,k+1} = \theta_{ij}^{l,k}-\eta \frac{\partial \mathcal{L}}{\partial \theta_{ij}^{l,k}}$, where $\frac{\partial \mathcal{L}}{\partial \theta_{i j}^{l,k}}=\beta\left(\sigma_{i j}^{l,k}\right)^{2}\left(\beta \Delta_{i j}^{l,k-1}+\mathcal{K}_j^{l+1}\left(\frac{1}{\sqrt{N_{l}}} h_{i}^{l}-\frac{\epsilon_{j}^{l+1}}{N_{l} v_{j}^{l+1}}\left(h_{i}^{l}\right)^{2} \mu_{i j}^{l}\right)\right)$, and $\Delta_{i j}^{l,k-1} = \theta_{ij}^{l,k-1} - \theta_{ij}^{l,k-2}$.}
		\end{algorithmic}
	\end{algorithm}

\section{Connection to elastic weight consolidation}
\label{proof}
In this section, we provide a proof of the elastic weight consolidation as a special example of VCL.
We first write $D\left(\boldsymbol{\theta}^k, \boldsymbol{\theta}^{k-1}\right)\equiv\mathcal{L}_{reg}=\mathrm{KL}\left[q_{\boldsymbol{\theta}^k}\left(\mathbf{w}\right) \| q_{\boldsymbol{\theta}^{k-1}}\left(\mathbf{w}\right)\right]$, and then we have
\begin{equation}
	D\left(\boldsymbol{\theta}^k, \boldsymbol{\theta}^{k-1}\right)=\int q(\mathbf{w} ; \boldsymbol{\theta}^k) \ln \frac{q(\mathbf{w} ; \boldsymbol{\theta}^k)}{q\left(\mathbf{w} ; \boldsymbol{\theta}^{k-1}\right)} d \mathbf{w},
\end{equation}
where the integral can be interpreted as the summation for the considered discrete weight variable.
Hereafter, we write $q_{\boldsymbol{\theta}^k}\left(\mathbf{w}\right)=q(\mathbf{w};\boldsymbol{\theta}^k)$.
We next assume the two consecutive solutions are sufficiently close, i.e.,
$\boldsymbol{\theta}^k\approx \boldsymbol{\theta}^{k-1}$, or $\bm{\theta}^k=\bm{\theta}^{k-1}+\Delta\bm{\theta}$ ($\Delta\bm{\theta}\to 0$), and then we
consider alternatively $D\left(\boldsymbol{\theta}^{k-1}, \boldsymbol{\theta}^{k}\right)$ because of small $\Delta\bm{\theta}$, and further
expand $\ln q(\mathbf{w};\boldsymbol{\theta}^k)$ around $\boldsymbol{\theta}^{k-1}$ up to the second order,
\begin{equation}
	\begin{aligned}
		D\left(\boldsymbol{\theta}^{k-1}, \boldsymbol{\theta}^{k}\right)& = \mathbb{E}_{q(\mathbf{w} ; \boldsymbol{\theta}^{k-1})}[\ln q(\mathbf{w} ; \boldsymbol{\theta}^{k-1}) - \ln q(\mathbf{w} ; \boldsymbol{\theta}^{k})],\\
		&\approx-\left(\boldsymbol{\theta}^{k}-\boldsymbol{\theta}^{k-1}\right)^{\top}\mathbb{E}_{q(\mathbf{w} ; \boldsymbol{\theta}^{k-1})}\left(\frac{\partial \ln q(\mathbf{w} ; \boldsymbol{\theta}^{k-1})}{\partial \boldsymbol{\theta}^{k-1}}\right)\\
		&-\frac{1}{2}\left(\boldsymbol{\theta}^{k}-\boldsymbol{\theta}^{k-1}\right)^{\top}\mathbb{E}_{q(\mathbf{w} ; \boldsymbol{\theta}^{k-1})}\left(\frac{\partial^{2}}{\partial (\boldsymbol{\theta}^{k-1})^2} \ln q(\mathbf{w} ; \boldsymbol{\theta}^{k-1})\right)\left(\boldsymbol{\theta}^{k}-\boldsymbol{\theta}^{k-1}\right).
	\end{aligned}
\end{equation}
We also find that
\begin{equation}
	\begin{aligned}
		\mathbb{E}_{q(\mathbf{w} ; \boldsymbol{\theta}^{k-1})}\frac{\partial \ln q(\mathbf{w} ; \boldsymbol{\theta}^{k-1})}{\partial \boldsymbol{\theta}^{k-1}} &= \int  q(\mathbf{w} ; \boldsymbol{\theta}^{k-1})\frac{1}{q(\mathbf{w} ; \boldsymbol{\theta}^{k-1})}\frac{\partial q(\mathbf{w} ; \boldsymbol{\theta}^{k-1})}{\partial \boldsymbol{\theta}^{k-1}} d\mathbf{w},\\
		& = \frac{\partial }{\partial \boldsymbol{\theta}^{k-1}}\int q(\mathbf{w} ; \boldsymbol{\theta}^{k-1}) d\mathbf{w},\\
		& = \frac{\partial 1}{\partial \boldsymbol{\theta}^{k-1}} = 0.
	\end{aligned}
\end{equation}
Notice that $\mathbb{E}_{q}\left(\frac{\partial^2\ln q}{\partial(\boldsymbol{\theta}^{k-1})^2}\right)=\mathbb{E}_{q}\left(\frac{1}{q}\frac{\partial^2q}{\partial(\bm{\theta}^{k-1})^2}-\Biggl(\frac{\partial\ln q}{\partial\boldsymbol{\theta}^{k-1}}\Biggr)^2\right)$,
where $q$ represents $q(\mathbf{w} ; \boldsymbol{\theta}^{k-1})$, and we have used $\frac{\partial^2\int d\mathbf{w}q}{\partial(\boldsymbol{\theta}^{k-1})^2}=0$, we finally arrive at
\begin{equation}
		D\left(\boldsymbol{\theta}^k, \boldsymbol{\theta}^{k-1}\right)= \frac{1}{2}\left(\boldsymbol{\theta}^{k}-\boldsymbol{\theta}^{k-1}\right)^{\top}F(\boldsymbol{\theta})\left(\boldsymbol{\theta}^{k}-\boldsymbol{\theta}^{k-1}\right),
\end{equation}
where $D(\boldsymbol{\theta}^k, \boldsymbol{\theta}^{k-1})\simeq D(\boldsymbol{\theta}^{k-1}, \boldsymbol{\theta}^{k})$ when $\Delta\bm{\theta}\to 0$,
and  $F(\boldsymbol{\theta})$ is exactly the Fisher information matrix whose definition is given by 
$F(\boldsymbol{\theta}) = \mathbb{E}_{q(\mathbf{w} ; \boldsymbol{\theta}^{k-1})}[\left(\frac{\partial}{\partial \boldsymbol{\theta}^{k-1}} \ln q(\mathbf{w} ; \boldsymbol{\theta}^{k-1})\right)\left(\frac{\partial}{\partial \boldsymbol{\theta}^{k-1}} \ln q(\mathbf{w} ; \boldsymbol{\theta}^{k-1})\right)^\top]$.
To conclude, when we take only the diagonal elements of the Fisher information matrix, we recover the elastic weight consolidation algorithm~\cite{PNAS-2017}.

\section{Details for replica computation}
\label{replica}
In this section, we demonstrate how to predict the generalization errors of variational continual learning by replica computation. 
First, we summarize our problem settings: we consider a teacher-student continual learning problem on binary perceptron,
where the student learns task 1 first and then task 2. For the data in these two tasks $\boldsymbol{x}^{1}$ and $\boldsymbol{x}^{2}$,
the labels are given by two teachers, $y^{1} = \operatorname{sign}(\sum_{i} W_i^{1}x_i^{1})$ and $y^{2} = \operatorname{sign}(\sum_{i} W_i^{2}x_i^2)$.
Note that, each dimension of the input data follows a uniform Bernoulli distribution, $x_i^{t,\mu} \in [-1,+1]$ and 
the training datasets of the two tasks consider different realizations, $\mathcal{D}_t = \{\boldsymbol{x}^{t,\mu},y^{t,\mu}\}_{\mu=1}^{M_t}$, where $t=1,2$. There is also a correlation between these two teachers, described by an overlap $r_0 = \frac{1}{N}\sum_i W_i^1 W_i^2$. Therefore, the joint distribution of teachers' weights can be parameterized as,
\begin{equation}
	P_0(W_i^1,W_i^2) =  \frac{1+r_0}{4}\delta(W_i^1-W_i^2) + \frac{1-r_0}{4}\delta(W_i^1+W_i^2),
\end{equation}
where we use the notation---teacher-average to denote the average over such distribution. For the sake of convenience,
we assume that during the learning of a certain task, the loss function is fixed. 
According to the variational theory in the main text, the loss functions for the continual learning are listed as follows,
\begin{equation}
	\begin{aligned}
		\mathcal{L}_1(\boldsymbol{m})&=-\sum_{\mu=1}^{M_1} \ln H\left(-\frac{\operatorname{sign}(\sum_{i} W_i^{1}x_i^{1,\mu})\sum_{i} m_{i} x_i^{1,\mu}}{\sqrt{\sum_{i}\left(1-m_{i}^{2}\right)}}\right),\\
		\mathcal{L}_2(\boldsymbol{m})&=-\sum_{\mu=1}^{M_2} \ln H\left(-\frac{\operatorname{sign}(\sum_{i} W_i^{1}x_i^{2,\mu})\sum_{i} m_{i} x_i^{2,\mu}}{\sqrt{\sum_{i}\left(1-m_{i}^{2}\right)}}\right) +\sum_{i=1}^N \mathrm{KL}(Q_{m_i} \|Q_{m_{1,i}}).
	\end{aligned}
\end{equation}
where $H(x)=\frac{1}{2}\mathrm{erfc}\left(\frac{x}{\sqrt{2}}\right)$ and $\boldsymbol{m}_{1}$ is the 
trained weight after learning task 1. In the following, the learning procedures of task 1 and task 2 will
be called as single-task learning and multi-task learning respectively, which actually reflects the learning's essence.

To predict the generalization errors in both learning scenarios, 
we apply replica method under the replica symmetry Ans\"atz. 
For a specific learning scenario, the derivations can be unfolded in two steps: First, we treat
the loss function for gradient-descent training as the Hamiltonian in canonical ensemble
and compute its averaged free energy, which entails the replica trick; Second, with the knowledge of the free energy, 
we show how to obtain the generalization errors of both tasks. In the following, the value of $M$ can be $M_1$ or $M_2$ depending on the learning stage.

\subsection{Thermodynamic system for single-task learning}
In this scenario, the thermodynamic system can be defined by a partition function,
\begin{equation}\label{Z-t1}
	Z = \int_{\Omega} \prod_{i=1}^N \mathrm{d} m_{i} e^{-\beta \mathcal{L}_1(\boldsymbol{m})}
	=\int_{\Omega^N} \prod_{i=1}^N \mathrm{d} m_{i} \prod_{\mu=1}^{M_1}  H^{\beta}\left(-\frac{\operatorname{sign}(\sum_{i} W_i^{1}x_i^{1,\mu})\sum_{i} m_{i} x_i^{1,\mu}}{\sqrt{\sum_{i}\left(1-m_{i}^{2}\right) }}\right),
\end{equation}
where $\Omega = [-1,+1]$. Our goal is to compute the quenched average of the free energy
$\langle \ln Z\rangle$ over the dataset $\mathcal{D}_{1}$ and teacher average. We can first remove
the average over $P(\boldsymbol{W}^{1})$ by performing a gauge 
transformation: $x_i^{1,\mu}\rightarrow W_i^{1}x_i^{1,\mu}$, $m_i\rightarrow W_i^{1}m_i$. 
The result can be seen as setting $W_i^{1}=1,\forall i$. 
Then, we apply replica trick, $\langle \ln Z\rangle = \lim_{n\to\infty}$ $\frac{\ln\left\langle Z^n\right\rangle }{n}$,
which requires to compute the replicated partition function,
\begin{equation}
	\langle Z^{n} \rangle=\int_{\Omega^{nN}} \prod_{a=1}^{n}   \prod_{i=1}^{N} \mathrm{d} m_{i}^{a} \left\langle\prod_{a=1}^{n}\prod_{\mu=1}^{M_1}  H^{\beta}\left(-\frac{\operatorname{sign}(\sum_{i} W_i^{1}x_i^{1,\mu})\sum_{i} m_{i}^a x_i^{1,\mu}}{\sqrt{\sum_{i}1-(m_{i}^a)^{2}}}\right)\right\rangle.
\end{equation}
Now, we introduce the local field,
\begin{equation}
	u^a = \frac{\sum_i m_i^a x_i^{1}}{\sqrt{N}},\qquad v_{1} =\frac{\sum_i W_i^{1} x_i^{1}}{\sqrt{N}},
\end{equation}
where the data index $\mu$ is omitted in advance. Note that, 
the statistics of local fields stem from the data distribution. 
According to central limit theorem, local fields should obey joint Gaussian distribution in
the thermodynamics limit $N\to \infty$. Thus, we have the following statistics
\begin{equation}
	\left\langle u^a \right\rangle = 0,\qquad \left\langle v_{1} \right\rangle = 0.
\end{equation}
In addition,
\begin{equation}
	\begin{aligned}
		\langle u^a u^a\rangle - \langle u^a \rangle \langle u^a \rangle &= \frac{\sum_i m_i^a m_i^a}{N},\\
		\langle u^a u^b\rangle - \langle u^a \rangle \langle u^b \rangle &= \frac{\sum_i m_i^a m_i^b}{N},\\
		\langle v_{1} u^a\rangle - \langle v_{1} \rangle \langle u^a \rangle &= \frac{\sum_i W_i^{1} m_i^a}{N},\\
	\end{aligned}
\end{equation}
where order parameters $q_{ab}=\frac{\sum_im_i^am_i^b}{N}$, $q_{aa}=\frac{\sum_im_i^a m_i^a}{N}$, $r_{a}^{1} = \frac{\sum_iW_i^{1}m_{i}^{a}}{N}$ naturally appear. 
We enforce the definitions of these order parameters to the replicated partition function by the Fourier Integral 
of Dirac delta functions, 
\begin{equation}
	\begin{aligned}
		\delta\left(\sum_i m_i^a m_i^b - q_{ab}N\right) &= \int \frac{1}{2\pi i}~e^{\hat{q}_{ab}\left(\sum_i^N m_i^a m_i^b - q_{ab}N\right)}d \hat{q}_{ab},\\
		\delta\left(\sum_i m_i^a m_i^a - q_{aa}N\right) &= \int \frac{1}{4\pi i }~e^{\frac{1}{2}\hat{q}_{aa}\left(\sum_i^N m_i^a m_i^a - q_{aa}N\right)}d \hat{q}_{aa},\\
		\delta\left(\sum_i W_i^{1} m_i^a - r_{a}^{1}N\right) &= \int \frac{1}{2\pi i}~e^{\hat{r}_{a}^{1}\left(\sum_i^N W_i^{1} m_i^a - r_{a}^{1}N\right)}d \hat{r}_{a}^{1},
	\end{aligned}
\end{equation}
and then we obtain,
\begin{equation}
	\begin{aligned}
		\langle Z^{n} \rangle &= \int \prod_a\frac{\mathrm{d}\hat{r}_{a}^{1} \mathrm{d}r_{a}^{1}}{2\pi i/N} \prod_{a} \frac{\mathrm{d} \hat{q}_{aa} \mathrm{d}q_{aa}}{4 \pi i/N} \prod_{a<b}  \frac{\mathrm{d} \hat{q}_{a b} \mathrm{d} q_{a b}}{2 \pi i/N} ~ e^{-N\sum_{a<b}\hat q_{ab}q_{ab}-\frac{1}{2}N\sum_a \hat q_{aa}q_{aa}-N\sum_a \hat{r}_{a}^{1} r_{a}^{1}}\\
		&~~\int_{\Omega^n}\prod_{a=1}^{n}\prod_{i=1}^N \mathrm{d} m_{i}^{a}  e^{\sum_{a<b} \hat{q}_{ab}\sum_i m_i^a m_i^b +\frac{1}{2}\sum_a \hat{q}_{aa}\sum_i m_i^a m_i^a} e^{\sum_a \hat{r}_{a}^{1}\sum_i m_i^a}\left\langle \prod_{\mu=1}^{M_1} \prod_{a=1}^{n} H^{\beta}\left(-\frac{\operatorname{sign}(v_{1})u^a}{\sqrt{ 1-q_{aa}}}\right)\right\rangle\\
	\end{aligned}
\end{equation}
Note that we introduce a prefactor $1/\sqrt{N}$ in the summations in Eq.~\eqref{Z-t1}, which does not affect the result.
Here, we consider the replica symmetry Ans\"atz: $q_{ab}=q_0, ~ \hat q_{ab} = \hat q_0,~q_{aa}=q_d,~\hat q_{aa}=\hat q_d, ~r_{a}^{1}=r_{1}, ~\hat{r}_{a}^{1}=\hat{r}_{1}$. Next, we will define and compute three terms separately and put them together in the final expression of the free energy:

The first term is the interaction term $G_{\mathrm{I}}$,
\begin{equation}
	\begin{aligned}
		G_{\mathrm{I}} &= -\frac{1}{2} \sum_{a, b} \hat{q}_{a b} q_{a b} - \sum_a \hat{r}_{a}^1r_{a}^1 \\
		&= -\frac{1}{2}\left(\sum_a \hat q_{aa} q_{aa} + \sum_{a\neq b} \hat q_{ab} q_{ab} \right) - n\hat{r}_1 r_1\\
		&= -\frac{1}{2}\left(n \hat q_d q_d + n(n-1) \hat q_{0} q_{0} \right) - n\hat{r}_1 r_1.\\
	\end{aligned}
\end{equation}

The second contribution is the entropy term $G_{\mathrm{S}}$,
\begin{equation}
	\begin{aligned}
		G_{\mathrm{S}}
		&=   \int_{[-1,1]^{n}} \prod_{a} \mathrm{d} m_{a} e^{\frac{1}{2} \sum_{a b} \hat{q}_{a b} m_{a} m_{b} + \hat r_1 \sum_a m_{a}  }\ ,\\
		&=   \int_{[-1,1]^{n}} \prod_{a} \mathrm{d} m_{a} e^{\frac{1}{2} \hat{q}_{d}\sum_{a}  m_{a} m_{a} - \frac{1}{2}\hat q_0 \sum_a m_a m_a + \frac{1}{2}\hat q_0(\sum_{a} m_a)^2 + \hat r_1 \sum_a m_{a} },\\
		&=  \int_{[-1,1]^{n}} \prod_{a} \mathrm{d} m_{a} e^{\frac{1}{2} \hat{q}_{d}\sum_{a}  m_{a} m_{a} - \frac{1}{2}\hat q_0 \sum_a m_a m_a + \hat r_1 \sum_a m_{a} }\int\mathcal{D}z ~ e^{\sqrt{\hat q_0}\sum_a m_a z},\\
		&=  \int\mathcal{D}z \left(\int_{-1}^{+1}  \mathrm{d} m ~e^{\frac{1}{2} \hat{q}_{d} m^2 -  \frac{1}{2}\hat q_0 m^2 +\sqrt{\hat q_0} m z+\hat r_1 m} \right)^n.\\
	\end{aligned}
\end{equation}

Finally, we compute the energy term $G_{\mathrm{E}}$,
\begin{equation}
	G_{\mathrm{E}} =  \left\langle  \prod_{a} H^{\beta}\left(-\frac{\operatorname{sign}({v_{1}})u^{a}}{\sqrt{ 1-q_d}}\right)\right\rangle,
\end{equation}
where $\left\langle \cdot \right\rangle $ denotes the average over the joint distribution of the local fields $(u^{a}$, $v_1)$.
Based on their statistics,
$\left\langle u^a u^a\right\rangle=  q_d$, $\left\langle u^a u^b\right\rangle=  q_{0}$, $\langle v_1
u^a\rangle =  r_{1}$, $\langle v_{1} v_{1}\rangle = 1$, they can thus be parametrized as
\begin{equation}\label{eq13}
	\begin{aligned}
		u^a &= \sqrt{q_d - q_0} \sigma_a + \sqrt{q_0} z,\\
		v_{1} &= \frac{ r_{1}}{\sqrt{q_0}} z +  \sqrt{1-\frac{r_{1}^2}{q_0}}y,
	\end{aligned}
\end{equation}
where $\sigma_a$, $z$, $y$ are all standard Gaussian variables. Substituting Eq.~\eqref{eq13} into the energy term arrives at
\begin{equation}\label{Energy-term-t1}
	\begin{aligned}
		G_{\mathrm{E}} &= \left\langle  \prod_{a} H^{\beta}\left(-\frac{\operatorname{sign}({v_{1}})u^{a}}{\sqrt{1-q_d}}\right) \right\rangle \\
		& = \int\mathcal{D}z\int \mathcal{D}y \prod_a \int\mathcal{D}\sigma_{a}~H^{\beta}\left(-\frac{\operatorname{sign}\left( \frac{ r_{1}}{\sqrt{q_0}} z +  \sqrt{1-\frac{r_{1}^2}{q_0}}y\right)\left(\sqrt{q_d-q_0} \sigma_{a} +  \sqrt{q_0}z\right)}{\sqrt{1-q_d}}\right)\\
		& =\int\mathcal{D}z\int\mathcal{D}y\left(\int\mathcal{D}\sigma~H^{\beta}\left(-\frac{\operatorname{sign}\left( \frac{ r_{1}}{\sqrt{q_0}} z +  \sqrt{1-\frac{r_{1}^2}{q_0}}y\right)\left(\sqrt{q_d-q_0} \sigma + \sqrt{q_0}z\right)}{\sqrt{1-q_d}}\right)\right)^n\\
		& = \int \mathcal{D}z\Biggl[H\left(-\frac{r_{1}}{\sqrt{q_0-r_{1}^2}}z\right)\left( \int\mathcal{D}\sigma~H^{\beta}\left(-\frac{\sqrt{q_d-q_0} \sigma + \sqrt{q_0}z}{\sqrt{1-q_d}}\right)\right)^n + \\
		& ~~~~ ~~~~ ~~~~ ~~~~ \left(1-H\left(-\frac{r_{1}}{\sqrt{q_0-r_{1}^2}}z\right)\right)\left( \int\mathcal{D}\sigma~H^{\beta}\left(\frac{\sqrt{q_d-q_0}\sigma + \sqrt{q_0}z}{\sqrt{1-q_d}}\right)\right)^n\Biggl]\\
		& = \int \mathcal{D}z~2H\left(-\frac{r_{1}}{\sqrt{q_0-r_{1}^2}}z\right)\left( \int\mathcal{D}\sigma~H^{\beta}\left(-\frac{\sqrt{q_d-q_0} \sigma + \sqrt{q_0}z}{\sqrt{1-q_d}}\right)\right)^n.\\
	\end{aligned}
\end{equation}
Note that $\mathcal{D}z$ indicates the standard Gaussian measure.

Finally, under the replica symmetry Ans\"atz, the replicated partition function can be written as
\begin{equation}
	\langle Z^{n} \rangle = \int \prod_a\frac{\mathrm{d}\hat{r}_{a}^{1} \mathrm{d}r_{a}^{1}}{2\pi i/N} \prod_{a} \frac{\mathrm{d} \hat{q}_{aa} \mathrm{d}q_{aa}}{4 \pi i/N} \prod_{a<b}  \frac{\mathrm{d} \hat{q}_{a b} \mathrm{d} q_{a b}}{2 \pi i/N} ~ e^{-Nnf_{\mathrm{RS}}}.
\end{equation}
Then, under the saddle-point approximation in the large $N$ limit, the free energy density is given by
\begin{equation}
	-\beta f_{\mathrm{RS}} = \lim_{n \rightarrow 0, N\rightarrow \infty} \frac{\ln\langle Z^n\rangle }{nN} = \lim_{n \rightarrow 0} 
	-\frac{1}{2}\left( \hat q_d q_d + (n-1) \hat q_{0} q_{0} \right) - \hat r_{1} r_{1} + \frac{\ln G_{\mathrm{S}}}{n} + \alpha_1 \frac{\ln G_{\mathrm{E}}}{n}.
\end{equation}
The free energy should be optimized with respect to the order parameters, and thus we have to derive the corresponding saddle-point equations through setting the gradients zero.
 We first compute $g_{\mathrm{S}} = \lim_{n \rightarrow 0}\frac{\ln G_{\mathrm{S}}}{n}$ and $g_{\mathrm{E}} =\lim_{n \rightarrow 0}\frac{\ln G_{\mathrm{E}}}{n}$,
\begin{equation}
	\begin{aligned}
		g_{\mathrm{S}}=\lim_{n \rightarrow 0} \frac{\ln G_{\mathrm{S}}}{n}
		&=\lim_{n \rightarrow 0} \frac{1}{n}  \int\mathcal{D}z \left(\int_{-1}^{+1}  \mathrm{d} m ~e^{\frac{1}{2} \hat{q}_{d} m^2 -  \frac{1}{2}\hat q_0 m^2 +\sqrt{\hat q_0} m z+\hat r_{1} m} \right)^n \\
		&=\int\mathcal{D}z\ln \left(\int_{-1}^{+1}  \mathrm{d} m ~e^{\frac{1}{2} \hat{q}_{d} m^2 -  \frac{1}{2}\hat q_0 m^2 +\sqrt{\hat q_0} m z+\hat r_{1} m} \right) .\\
		g_{\mathrm{E}}=\lim_{n \rightarrow 0}\frac{\ln G_{\mathrm{E}}}{n}
		&= \ln \int \mathcal{D}z~2H\left(-\frac{r_{1}}{\sqrt{q_0-r_{1}^2}}z\right)\left( \int\mathcal{D}\sigma~H^{\beta}\left(-\frac{\sqrt{q_d-q_0} \sigma + \sqrt{q_0}z}{\sqrt{1-q_d}}\right)\right)^n\\
		& = \int \mathcal{D}z~2H\left(-\frac{r_{1}}{\sqrt{q_0-r_{1}^2}}z\right)\ln \int\mathcal{D}\sigma~H^{\beta}\left(-\frac{\sqrt{q_d-q_0} \sigma + \sqrt{q_0}z}{\sqrt{1-q_d}}\right).
	\end{aligned}
\end{equation}
Thus, the saddle-point equations can be expressed as,
\begin{equation}\label{sd-t1}
	\begin{aligned}
		q_{d} &= 2\frac{\partial g_{S} }{\partial \hat{q}_{d}}, & q_{0}& = -2\frac{\partial g_{S}}{\partial \hat{q}_{0}},  & r_{1} &= \frac{\partial g_{S}}{\partial \hat{r}_{1}};\\
		\hat{q}_{d} &= 2\alpha_1\frac{\partial g_{E}}{\partial q_{d}}, & \hat{q}_{0} &= -2\alpha_1\frac{\partial g_{E}}{\partial q_{0}}, &  \hat{r}_{1} &= \alpha_1\frac{\partial g_{E}}{\partial r_{1}}.
	\end{aligned}
\end{equation}
To make the expressions more compact, we define a probability measure,
\begin{equation}
	\left\langle \left\langle \mathcal{O} \right\rangle \right\rangle_{\mathcal{H}(m)} =  \frac{\int_{-1}^{+1}\mathcal{O}~ e^{\mathcal{H}(m)} \mathrm{d} m  }{\int_{-1}^{+1} e^{\mathcal{H}(m)} \mathrm{d} m }.
\end{equation}
Then, we have,
\begin{equation}\label{sd-task1-gs}
	\begin{aligned}
		q_{d} 
		&= 2\int \mathcal{D} z \frac{ \int_{-1}^{+1} \frac{1}{2}m^2 ~e^{\mathcal{I}(m,z)} \mathrm{d} m }{ \int_{-1}^{+1} e^{\mathcal{I}(m,z)} \mathrm{d} m } = \int \mathcal{D} z \left\langle \left\langle m^{2} \right\rangle \right\rangle_{\mathcal{I}(m,z)},\\
		q_{0}
		&= - 2\int \mathcal{D} z \frac{ \int_{-1}^{+1} \left(-\frac{1}{2}m^2 + \frac{z}{2\sqrt{\hat{q}_{0}}} m \right)  ~e^{\mathcal{I}(m,z)} \mathrm{d} m }{ \int_{-1}^{+1} e^{\mathcal{I}(m,z)} \mathrm{d} m }= \int \mathcal{D} z \left\langle \left\langle m^{2} - \frac{z}{\sqrt{\hat{q}_0}}m \right\rangle \right\rangle_{\mathcal{I}(m,z)},\\
		r_{1}
		&= \int \mathcal{D} z \frac{ \int_{-1}^{+1} m ~ e^{\mathcal{I}(m,z)} \mathrm{d} m }{ \int_{-1}^{+1} e^{\mathcal{I}(m,z)} \mathrm{d} m  } =  \int \mathcal{D} z \left\langle \left\langle m  \right\rangle \right\rangle_{\mathcal{I}(m,z)},\\
	\end{aligned}
\end{equation}
where
\begin{equation}
	\mathcal{I} (m,z) =\frac{1}{2} \hat{q}_{d} m^2 -  \frac{1}{2}\hat q_0 m^2 +\sqrt{\hat q_0} m z+\hat r_{1} m.
\end{equation}
Th other derivatives related to $g_{\mathrm{E}}$ can be computed as,
\begin{equation}\label{sd-task1-ge}
	\begin{aligned}
		\hat{q}_{d} &= 4\alpha_1\int \mathcal{D} z~H\left(\lambda(z) \right) \frac{\int\mathcal{D}\sigma~\beta H^{\beta-1}\left(\gamma(z,\sigma)\right)H^{\prime}\left(\gamma(z,\sigma)\right) u(z,\sigma) }{\int\mathcal{D}\sigma~H^{\beta}\left(\gamma(z,\sigma)\right)},\\
		\hat{q}_{0} &= -4\alpha_1 \int \mathcal{D} z ~ H^{\prime}\left(\lambda(z) \right) v_{1}(z) \ln \int\mathcal{D}\sigma~H^{\beta}\left(\gamma(z,\sigma)\right)\\
		&\quad -4\alpha_1 \int \mathcal{D} z ~ H\left(\lambda(z) \right) \frac{\int\mathcal{D}\sigma~\beta H^{\beta-1}\left(\gamma(z,\sigma)\right)H^{\prime}\left(\gamma(z,\sigma)\right)w(z,\sigma) }{\int\mathcal{D}\sigma~H^{\beta}\left(\gamma(z,\sigma)\right)},\\
		\hat{r}_{1} &= 2\alpha_1 \int \mathcal{D} z ~ H^{\prime}\left(\lambda(z) \right) h(z) \ln \int\mathcal{D}\sigma~H^{\beta}\left(\gamma(z,\sigma)\right),
	\end{aligned}
\end{equation}
where
\begin{equation}\label{sd-task1-ge-where}
	\begin{aligned}
		\lambda(z) &= -\frac{r_{1}}{\sqrt{q_0-r_{1}^2}}z, \\
		\gamma(z,\sigma) &= -\frac{\sqrt{q_d-q_0} \sigma + \sqrt{q_0}z}{\sqrt{1-q_d}},\\
		u(z,\sigma) & =  \frac{(q_{0}-1)\sigma -\sqrt{q_{0}}\sqrt{q_{d}-q_{0}}z}{2(1-q_{d})^{\frac{3}{2}}\sqrt{q_{d}-q_{0}} },\\
		v_{1}( z ) &= \frac{r_{1} z}{2(q_{0}-r_{1}^{2})^{\frac{3}{2}}},\\
		w(z,\sigma) &=  \frac{\sqrt{q_{0}}\sigma-\sqrt{q_{d}-q_{0}}z }{2\sqrt{1-q_{d}}\sqrt{q_{d}-q_{0}}\sqrt{q_{0}} },\\
		h(z) &= \frac{-q_{0}z}{(q_{0}-r_{1}^{2})^{\frac{3}{2}}}.
	\end{aligned}
\end{equation}

\subsubsection{Generalization error for task 1}

During the learning of task 1, the generalization error can be defined as
\begin{equation}
	\epsilon_{g}^{1} = \left\langle \Theta \left(-\operatorname{sign}\Biggl(\sum_i W_i^{1} x^*_i\Biggr) \sum_{i} \operatorname{sign}(m_i) x^*_{i} \right) \right\rangle,
\end{equation}
where $\boldsymbol{x}^{*}$ is one fresh example (or a test data). Note that, the average $\langle \cdot \rangle$ 
refers to the ensemble average based on the thermodynamic system for task 1.
To handle this average, we define similar local fields for test data,
\begin{equation}
	u^* = \frac{\sum_i \operatorname{sign}(m_i) x_i^*}{\sqrt{N}}, \quad v_{1}^* =\frac{\sum_i W_i^{1} x_i^*}{\sqrt{N}},
\end{equation}
whose statistical properties are as follows, $\langle v_{1}^* v_{1}^*\rangle=1$, $\langle u^{*} u^{*} \rangle=1$, $p_{1} =\langle v_{1}^* u^{*} \rangle = \frac{1}{N}\sum_{i} \operatorname{sign}(m_i) W_i^{1} $. 
Then, local fields are parametrized as
\begin{equation}
	\begin{aligned}
		u^* &= z,\\
		v_{1}^* &= p_{1} z + \sqrt{1-p_{1}^{2}}y.
	\end{aligned}
\end{equation} 
The generalization error becomes
\begin{equation}
	\begin{aligned}
		\epsilon_{g}^{1}
		&= \int \mathcal{D}z\int \mathcal{D} y ~ \Theta \left(-\operatorname{sign}(p_{1} z + \sqrt{1-p_{1}^{2}}y ) z \right)\\
		&= \int \mathcal{D}z ~ 2H\left(-\frac{p_{1}}{\sqrt{1-p_{1}^2}}z\right)\Theta(-z)\\
		&=  \frac{1}{\pi}\arccos (p_1),
	\end{aligned}
\end{equation}
which introduces a new order parameter $p_1$. To obtain the value of $p_1$, we should first introduce the order parameter
to the original replicated partition function by a Fourier integral representation of the Dirac delta function. 
After some manipulations, the free energy density under the replica symmetry Ans\"atz and the limit of $n\to 0$ is rewritten as
\begin{equation}
	-\beta f_{\mathrm{RS}} = -\frac{1}{2}\left( \hat q_d q_d + (n-1) \hat q_{0} q_{0} \right) - \hat r_{1} r_{1} - \hat p_{1} p_{1} + g^{\prime}_{\mathrm{S}} + \alpha_1 g_{\mathrm{E}}
\end{equation} 
where the energy term $g_{\mathrm{E}}$ remains the same, but the entropy term $g^{\prime}_{\mathrm{S}}$ changes as follows,
\begin{equation}
	g_{S}'= \int\mathcal{D}z\ln \left(\int_{-1}^{+1}  \mathrm{d} m ~e^{\frac{1}{2} \hat{q}_{d} m^2 -  \frac{1}{2}\hat q_0 m^2 +\sqrt{\hat q_0} m z+\hat r_{1} m +\hat p_{1} \operatorname{sign}(m)} \right).
\end{equation}
In the new saddle-point equations, it is easy to find that $\hat{p}_{1}=\alpha_1\frac{\partial g_{E}}{\partial p_{1}}=0$,
which means the original saddle-point equations [Eqs.~\eqref{sd-task1-gs},~\eqref{sd-task1-ge}] are independent of the new order parameters $p_1$ and $\hat{p}_1$. 
Thus, after the convergence of iterating the original saddle-point equations, 
we compute $p_1$ by ${p}_{1}=\frac{\partial g^{\prime}_{S}}{\partial \hat{p}_{1}}$ and obtain the following result as
\begin{equation}
	p_{1}  =  \int \mathcal{D} z \left\langle \left\langle \operatorname{sign}(m)  \right\rangle \right\rangle_{\mathcal{I}(m,z)}.
\end{equation}

\subsubsection{Generalization error for task 2}

The calculation of generalization error for task 2 follows a similar line, except for one important difference,
which is the involvement of the teacher-average over the joint distribution $P(\boldsymbol{W}^1,\boldsymbol{W}^2)$. 
In the following, we will omit the similar process in computing $\epsilon_{g}^{1}$, and instead focus on the 
treatment of teacher-average.

In analogous to $\epsilon_{g}^{1}$, the generalization error for task 2 can be expressed as
\begin{equation}
	\epsilon_{g}^{2} =  \frac{1}{\pi}\arccos (p_2),
\end{equation}
where $p_{2}  = \frac{1}{N}\sum_{i} \operatorname{sign}(m_i) W_i^{2} $. Introducing $p_2$ to the replicated partition
function results in a modified entropy term before taking the replica symmetry Ans\"atz,
\begin{equation}
	\begin{aligned}
		(G^{\prime}_{\mathrm{S}})^{N} 
		&= \int_{\Omega^{nN}} \prod_{a=1}^{n}\prod_{i=1}^N \mathrm{d} m_{i}^{a}   e^{\sum_{a<b} \hat{q}_{ab}\sum_i m_i^a m_i^b +\frac{1}{2}\sum_a \hat{q}_{aa}\sum_i m_i^a m_i^a } \mathbb{E}_{\mathrm{T}} \left[e^{ \sum_a \hat r_{a}^{1}\sum_i W_i^{1} m_i^a+ \sum_a \hat{p}_{a}^{2} \sum_i  W_i^{2} \operatorname{sign}(m_i^a)} \right]\\
		&= \int_{\Omega^{nN}} \prod_{a=1}^{n}\prod_{i=1}^N \mathrm{d} m_{i}^{a}   e^{\sum_{a<b} \hat{q}_{ab}\sum_i m_i^a m_i^b +\frac{1}{2}\sum_a \hat{q}_{aa}\sum_i m_i^a m_i^a } \mathbb{E}_{\mathrm{T}} \left[e^{ \sum_a \hat r_{a}^{1}\sum_i m_i^a+ \sum_a \hat{p}_{a}^{2} \sum_i  W_i^{1} W_i^{2} \operatorname{sign}(m_i^a)} \right],\\
	\end{aligned}
\end{equation}
where the $\mathbb{E}_{\mathrm{T}}[\cdot]$ denotes the teacher average.
A gauge transformation $m_i^a\to m_i^a  W_i^{1}$ is used in the second equality. Thus, the expectation can be computed as,
\begin{equation}\label{teacher-average-t1}
	\begin{aligned}
		& \mathbb{E}_{\mathrm{T}}\left[e^{ \sum_a \hat r_{a}^{1}\sum_i m_i^a+ \sum_a \hat{p}_{a}^{2} \sum_i  W_i^{1} W_i^{2} \operatorname{sign}(m_i^a)} \right]\\
		& = \sum_{\boldsymbol{W}^{1},\boldsymbol{W}^{2}}\prod_{i=1}^N P_0(W_i^{1},W_i^{2}) ~e^{ \sum_a \hat r_{a}^{1}\sum_i m_i^a+ \sum_a \hat{p}_{a}^{2} \sum_i  W_i^{1} W_i^{2} \operatorname{sign}(m_i^a)}\\
		& = \prod_{i=1}^N \sum_{W_i^{1},W_i^{2}}  P_0(W_i^{1},W_i^{2}) ~e^{ \sum_a \hat r_{a}^{1} m_i^a+ \sum_a \hat{p}_{a}^{2}   W_i^{1} W_i^{2} \operatorname{sign}(m_i^a)}\\
		& = \prod_{i=1}^N \Biggl[ \frac{1+r_0}{2} \cosh\left( \sum_a \hat r_{a}^{1} m_i^a +\sum_a \hat p_{a}^{2} \operatorname{sign}(m_i^a)  \right) + \frac{1- r_0}{2} \cosh\left( \sum_a \hat r_{a}^{1} m_i^a -\sum_a \hat p_{a}^{2} \operatorname{sign}(m_i^a)  \right)   \Biggr].
	\end{aligned}
\end{equation}
Under the replica symmetry Ans\"atz, we entropy term becomes,
\begin{equation}
	\begin{aligned}
		G^{\prime}_{\mathrm{S}}
		&=  \int\mathcal{D}z \Biggl[\left(\int_{-1}^{+1}  \mathrm{d} m ~e^{\frac{1}{2} \hat{q}_{d} m^2 -  \frac{1}{2}\hat q_0 m^2 +\sqrt{\hat q_0} m z+\hat{r}_{1} m + \hat{p}_{2} \operatorname{sign}(m)} \right)^n \frac{1+r_0}{2}\\
		& \quad + \left(\int_{-1}^{+1}  \mathrm{d} m ~e^{\frac{1}{2} \hat{q}_{d} m^2 -  \frac{1}{2}\hat q_0 m^2 +\sqrt{\hat q_0} m z+\hat{r}_{1} m-\hat{p}_{2} \operatorname{sign}(m)} \right)^n \frac{1-r_0}{2}\Biggr].\\
	\end{aligned}
\end{equation}
After taking the limitation of $n\to 0$, we have,
\begin{equation}
	\begin{aligned}
		g^{\prime}_{\mathrm{S}} &=  \int\mathcal{D}z \ln \left(\int_{-1}^{+1}  \mathrm{d} m ~e^{\frac{1}{2} \hat{q}_{d} m^2 -  \frac{1}{2}\hat q_0 m^2 +\sqrt{\hat q_0} m z+\hat{r}_{1} m + \hat{p}_{2} \operatorname{sign}(m)} \right) \frac{1+r_0}{2}\\
		& \quad + \int\mathcal{D}z\ln \left(\int_{-1}^{+1}  \mathrm{d} m ~e^{\frac{1}{2} \hat{q}_{d} m^2 -  \frac{1}{2}\hat q_0 m^2 +\sqrt{\hat q_0} m z+\hat{r}_{1} m-\hat{p}_{2} \operatorname{sign}(m)} \right) \frac{1-r_0}{2}.\\
	\end{aligned}
\end{equation}
Therefore, we can finally get $p_2$ by ${p}_{2}=\frac{\partial g^{\prime}_{S}}{\partial \hat{p}_{2}}$, which results in
\begin{equation}
	p_{2}  =   r_0\int \mathcal{D} z \left\langle \left\langle \operatorname{sign}(m)  \right\rangle \right\rangle_{\mathcal{I}(m,z)} = r_0 p_{1},
\end{equation}
which is actually linear to $p_{1}$, and $r_0$ characterizes the task similarity.

\subsubsection{Constrained partition function for learning process}

The partition function Eq.~\eqref{Z-t1} defined previously focuses on the stationary state of the system, which helps to predict the optimal performance of learning given the corresponding Hamiltonian (loss function). 
However, by introducing a constraint in this system, we can study the stationary state during the learning process. The constrained partition function can be written as,
\begin{equation}\label{Z-constrain}
	Z = \int_{\Omega} \prod_{i=1}^N \mathrm{d} m_{i} \delta \left( \sum_i m_i^2 - q_{\star} N \right) e^{-\beta \mathcal{L}_1(\boldsymbol{m})},
\end{equation}
where $N-q_{\star}N$ is the total extent of the weight fluctuation. As the learning goes on,  $q_{\star}$ will gradually increase until a saturation to $1$. 
Therefore, to explore the stationary state during learning, we can set a value of $q_{\star}$ manually, and 
then solve the corresponding partition function Eq.~\eqref{Z-constrain}.     

Note that, the calculation of the constrained partition function follows the similar procedure for the non-constrained one,
except for one subtle difference, which is that the order parameter $q_{aa}$ as well as $q_d$ under the replica symmetry Ans\"atz
are replaced by a pre-specified constant $q_{\star}$. Hence, we can directly derive the free energy density $-\beta f_{\mathrm{RS}}$,
\begin{equation}
	-\beta f_{\mathrm{RS}} = \lim_{n \rightarrow 0} 
	-\frac{1}{2}\left( \hat q_{d} q_{\star} + (n-1) \hat q_{0} q_{0} \right) - \hat r_{1} r_{1} + \frac{\ln G_{\mathrm{S}}}{n} + \alpha_1 \frac{\ln G^{\star}_{\mathrm{E}}}{n},
\end{equation}
where the entropy term remains unchanged, and the energy term becomes
\begin{equation}
	g^{\star}_{\mathrm{E}}=\lim_{n \rightarrow 0}\frac{\ln G^{\star}_{\mathrm{E}}}{n}
	 = \int \mathcal{D}z~2H\left(-\frac{r_{1}}{\sqrt{q_0-r_{1}^2}}z\right)\ln \int\mathcal{D}\sigma~H^{\beta}\left(-\frac{\sqrt{q_{\star}-q_0} \sigma + \sqrt{q_0}z}{\sqrt{1-q_{\star}}}\right).
\end{equation}
Notice that $q_{\star}$ is a constant, and thus the saddle-point equations get simplified to
\begin{equation}\label{sd-t1-con}
	q_{\star} = 2\frac{\partial g_{S} }{\partial \hat{q}_{d}}, \quad q_{0} = -2\frac{\partial g_{S}}{\partial \hat{q}_{0}},  \quad r_{1} = \frac{\partial g_{S}}{\partial \hat{r}_{1}},\quad
	\hat{q}_{0} = -2\alpha_1\frac{\partial g^{\star}_{E}}{\partial q_{0}}, \quad \hat{r}_{1} = \alpha_1\frac{\partial g^{\star}_{E}}{\partial r_{1}},
\end{equation}
where the first one is an explicit equation for $\hat{q}_{d}$ whose value can be numerically found (e.g., by using the secant method).

\subsection{Thermodynamic system for multi-task learning}
In the scenario of multi-task learning, a distinct characteristic is that previous task information is
incorporated into the learning procedure of the current task by a regularization term. 
The task information from previous task refers to the trained weight for task 1, which
can be captured by the associated partition function in the single-task learning section. In terms of the current
task, a similar partition function should be established once the trained weights of the first task are given.
This observation indicates that the variational continual learning \textit{can be mapped} to
the form of Franz-Parisi potential originally proposed in spin glass theory~\cite{Franz-1995} and later in neural networks~\cite{Huang-2014},
\begin{equation}
	\Phi = \frac{1}{\tilde{Z}} \int_{\tilde{\Omega}}\prod_{i=1}^N \mathrm{d} \tilde{m}_{i}~e^{\tilde{\beta} \mathcal{L}_1(\tilde{\boldsymbol{m}})} \ln\int_{\Omega} \prod_{i=1}^N \mathrm{d} m_{i}~  e^{\beta \mathcal{L}_2(\boldsymbol{m},\tilde{\boldsymbol{m}})},
\end{equation}
where
\begin{equation}
	\begin{aligned}
		\mathcal{L}_1(\tilde{\boldsymbol{m}})&=\sum_{\mu=1}^{M_1} \ln H\left(-\frac{\operatorname{sign}(\sum_{i} W_i^{1}x_i^{1,\mu})\sum_{i} m_{i} x_i^{1,\mu}}{\sqrt{\sum_{i}\left(1-m_{i}^{2}\right)}}\right),\\
		\mathcal{L}_2(\boldsymbol{m},\tilde{\boldsymbol{m}})&=\sum_{\mu=1}^{M_2} \ln H\left(-\frac{\operatorname{sign}(\sum_{i} W_i^{1}x_i^{2,\mu})\sum_{i} m_{i} x_i^{2,\mu}}{\sqrt{\sum_{i}\left(1-m_{i}^{2}\right)}}\right) - \sum_{i=1}^N \mathrm{KL}(Q_{m_i}\|Q_{m_i^{1}}).
	\end{aligned}
\end{equation}
Next, our goal is to compute the quenched disorder average of the thermodynamic potential $\Phi$, where the averages are threefold, consisting of two data averages over $\mathcal{D}_1$ and $\mathcal{D}_2$, as well as the teacher average, and can be explicitly worked out as
\begin{equation}
	\begin{aligned}
		\left\langle \Phi \right\rangle
		&= \mathbb{E}_{\mathrm{T}} ~\left\langle \frac{1}{\tilde{Z}} \int_{\tilde{\Omega}}\prod_{i=1}^N \mathrm{d} \tilde{m}_{i}~e^{\tilde{\beta} \mathcal{L}_1(\tilde{\boldsymbol{m}})} \ln\int_{\Omega} \prod_{i}^N \mathrm{d} m_{i}~e^{\beta \mathcal{L}_2(\boldsymbol{m},\tilde{\boldsymbol{m}})}\right\rangle_{\mathcal{D}_1,\mathcal{D}_2}\\
		&= \mathbb{E}_{\mathrm{T}} ~\frac{1}{\tilde{Z}} \int_{\tilde{\Omega}}\prod_{i=1}^N \mathrm{d} \tilde{m}_{i}\prod_{\mu=1}^{M_1}  \left\langle H^{\tilde\beta}\left(-\frac{\operatorname{sign}(\sum_{i} W_i^{1}x_i^{1,\mu})\sum_{i} \tilde{m}_{i} x_i^{1,\mu}}{\sqrt{\sum_{i}\left(1-\tilde{m}_{i}^{2}\right) }}\right)\right\rangle_{\mathcal{D}_{1}}\\
		&~~~~\ln\int_{\Omega} \prod_{i=1}^N \mathrm{d} m_{i}~\prod_{\mu=1}^{M_2} \left\langle H^{\beta}\left(-\frac{\operatorname{sign}(\sum_{i} W_i^{2} x_i^{2,\mu})\sum_{i} m_{i} x_i^{2,\mu}}{\sqrt{\sum_{i}\left(1-m_{i}^{2}\right) }}\right)\right\rangle_{\mathcal{D}_{2}} e^{-\beta\sum_{i}\mathrm{KL}(m_i,\tilde{m}_i)  }.\\	
	\end{aligned}
\end{equation}
Note that two data averages decouple directly due to the independence between two datasets (but the labels can be correlated).
We now omit the subscripts for the data averages. To start the calculation, we introduce two useful replica formulas,
	$\frac{1}{\tilde{Z}} = \lim_{n \rightarrow 0} \tilde{Z}^{n-1}$, $\ln Z = \lim_{s \rightarrow 0} \partial_s Z^s.$
Then the potential turns out to be
\begin{equation}
	\begin{aligned}
		\left\langle \Phi \right\rangle
		&= \lim_{n \rightarrow 0} \lim_{s \rightarrow 0} \partial_s ~\mathbb{E}_{\mathrm{T}}~ \int_{\tilde{\Omega}^{nN}}\prod_{a=1}^{n}\prod_{i=1}^N \mathrm{d} \tilde{m}_{i}^a \prod_{a=1}^{n} \prod_{\mu=1}^{M_1}  \left\langle H^{\tilde{\beta}}\left(-\frac{\operatorname{sign}(\sum_{i}W_i^{1} x_i^{1,\mu})\sum_{i} \tilde{m}_{i}^{a} x_i^{1,\mu}}{\sqrt{\sum_{i}\left(1-(\tilde{m}_{i}^{a})^{2}\right) }}\right)\right\rangle\\
		&~~~~\int_{\Omega^{sN}} \prod_{c=1}^{s}\prod_{i=1}^N \mathrm{d} m_{i}^{c}~ \prod_{c=1}^{s}\prod_{\mu=1}^{M_2} \left\langle H^{\beta}\left(-\frac{\operatorname{sign}(\sum_{i} W_i^{2} x_i^{2,\mu})\sum_{i} m_{i}^{c} x_i^{2,\mu}}{\sqrt{\sum_{i}\left(1-(m_{i}^{c})^{2}\right) }}\right)\right\rangle e^{-\beta\sum_c\sum_{i}\mathrm{KL}(m_i^{c}, \tilde{m}_i^{a=1})  }.\\	
	\end{aligned}
\end{equation}

Local fields are introduced for both loss functions,
\begin{equation}
	\tilde{u}^a = \frac{\sum_i \tilde{m}_i^a x_i^{1}}{\sqrt{N}},\quad \tilde{v}_{1} =\frac{\sum_i W_i^{1} x_i^{1}}{\sqrt{N}}, \quad u^c = \frac{\sum_i m_i^c x_i^{2}}{\sqrt{N}},\quad v_{2} =\frac{\sum_i W_i^{2} x_i^{2}}{\sqrt{N}}, 
\end{equation}
where we omit superscript $\mu$. Based on the central limit theorem, 
the local fields follow the joint Gaussian distribution with zero mean 
and the non-zero second moments as,
\begin{equation}
	\begin{aligned}
		\langle \tilde{u}^a \tilde{u}^a\rangle&= \frac{\sum_i \tilde m_i^a \tilde m_i^a}{N},&
		\langle \tilde{u}^a \tilde{u}^b\rangle&= \frac{\sum_i \tilde m_i^a \tilde m_i^b}{N},&
		\langle \tilde{v}_{1} \tilde{u}^a\rangle&= \frac{\sum_i W_i^{1} \tilde m_i^a}{N},&
		\langle\tilde{v}_{1} \tilde{v}_{1} \rangle&= 1,\\
		\langle u^c u^c\rangle &= \frac{\sum_i m_i^c m_i^c}{N},&
		\langle u^c u^d\rangle &= \frac{\sum_i m_i^c m_i^d}{N},&
		\langle v_{2} u^c\rangle &= \frac{\sum_i W_i^{2} m_i^c}{N},&
		\langle v_{2} v_{2}\rangle &= 1.\\
	\end{aligned}
\end{equation}
We can therefore define the order parameters, 
$\tilde q_{aa}=\frac{\sum_i \tilde m_i^a \tilde m_i^a}{N}$, $\tilde q_{ab}=\frac{\sum_i \tilde m_i^a \tilde m_i^b}{N}$, $\tilde{r}_{a}^{1} = \frac{\sum_i W_i^{1} \tilde m_i^a}{N}$, $q_{cc}=\frac{\sum_im_i^cm_i^c}{N}$, $q_{cd}=\frac{\sum_im_i^c m_i^d}{N}$, $r_{c}^{2} = \frac{\sum_{i}W_i^{2}m_{i}^{c}}{N}$
and enforce these definitions in potential $\Phi$ by Dirac delta function $\delta(\cdot)$. 
After a few algebra manipulations, we arrive at
\begin{equation}
	\left\langle \Phi \right\rangle=\lim_{n \rightarrow 0} \lim_{s \rightarrow 0} \partial_s\int 
	\prod_{a}\frac{\mathrm{d}\hat{\tilde q}_{aa} \mathrm{d}\tilde q_{aa}}{4 \pi i/N}
	\prod_{a<b}\frac{\mathrm{d} \hat{\tilde q}_{a b} \mathrm{d} \tilde q_{a b}}{2 \pi i/N} \prod_a\frac{\mathrm{d}\hat{\tilde r}_{a}^{1}\mathrm{d}\tilde{r}_{a}^1}{2\pi i/N}\prod_{c} \frac{\mathrm{d} \hat{q}_{cc} \mathrm{d}q_{cc}}{4 \pi i/N}\prod_{c<d}  \frac{\mathrm{d} \hat{q}_{cd} \mathrm{d} q_{cd}}{2 \pi i/N}
	\prod_c\frac{\mathrm{d}\hat{r}_{c}^{2} \mathrm{d}r_{c}^{2}}{2\pi i/N}
	e^{N\mathcal{S}},
\end{equation}
where a similar manipulation of the teacher average to Eq.~\eqref{teacher-average-t1} is carried out, and the action $\mathcal{S}$ finally reads  
\begin{equation}
	\begin{aligned}
		\mathcal{S} &= -\frac{1}{2} \sum_{a, b} \hat{\tilde q}_{a b} \tilde q_{a b} - \sum_a \hat{\tilde r}_{a}^{1} \tilde{r}_{a}^{1} -\frac{1}{2} \sum_{c, d} \hat{q}_{cd} q_{cd} - \sum_c \hat r_{c}^{2} r_{c}^{2}\\
		&\quad + \ln\int_{\tilde{\Omega}^{nN}} \int_{{\Omega}^{sN}} \prod_{a} \mathrm{d} \tilde{m}_{a} \prod_{c}\mathrm{d} m_{c} ~ e^{\frac{1}{2}\sum_{a,b} \hat{\tilde q}_{ab} \tilde m_a\tilde m_b +  +\frac{1}{2}\sum_{c,d} \hat{q}_{cd} m_c m_d-\beta\sum_c \mathrm{KL}(m_i^c, 	 \tilde m_i^{a=1}) }\\
		&\quad\times \left[ \frac{1+r_0}{2} \cosh\left( \sum_a \hat{\tilde{r}}_{a}^{1} \tilde{m}_i^a +\sum_c \hat{r}_{c}^{2} m_i^c  \right) +  \frac{1- r_0}{2} \cosh\left( \sum_a \hat{\tilde{r}}_{a}^{1} \tilde{m}_i^a -\sum_c \hat{r}_{c}^{2} m_i^c  \right)   \right] \\
		&\quad+\alpha_1\ln\left\langle \prod_{a=1}^{n} H^{\tilde\beta}\left(-\frac{\operatorname{sign}({\tilde{v}_{1}}) \tilde u^{a}}{\sqrt{ 1-\tilde{q}_{aa}}}\right)\right\rangle + \alpha_2 \ln \left\langle \prod_{c=1}^{s} H^{\beta}\left(-\frac{\operatorname{sign}({v_{2}}) u^{c}}{\sqrt{ 1-q_{cc}}}\right)\right\rangle.
	\end{aligned}
\end{equation}
The maximum of $\mathcal{S}$ dominates the integrand under the large $N$ limit.
Thus, we derive the saddle-point equations by taking derivatives of the action $\mathcal{S}$ with respect to the order parameters.
We then apply the replica symmetry Ans\"atz,
\begin{equation}
	\begin{aligned}
		\tilde q_{ab}&=\tilde q_0,
		 &\hat{\tilde q}_{ab} &= \hat{\tilde q}_0,
		 &\tilde q_{aa}&=\tilde q_d,
		 &\hat{\tilde q}_{aa}&=\hat{\tilde q}_d,
		 &\tilde{r}_{a}^{1}&=\tilde{r}_1, 
		 &\hat{\tilde r}_{a}^{1}&= \hat{\tilde{r}}_1,\\
		 q_{cd}&=q_0, 
		 & \hat q_{cd} &= \hat q_0, 
		 & q_{cc} &= q_d, 
		 &\hat q_{cc}&=\hat q_d, 
		 & r_{c}^{2}&=r_{2}, 
		 &\hat{r}_{c}^{2}&=\hat{r}_{2}.
	\end{aligned}
\end{equation}

To make the calculation neat, we divide the action into three parts and compute their contributions respectively. 
First, the interaction term reads
\begin{equation}
	\begin{aligned}
		\mathcal{G}_{\mathrm{I}}
		&=-\frac{1}{2} \sum_{a, b} \hat{\tilde q}_{a b} \tilde q_{a b} - \sum_a \hat{\tilde r}_{a}^{1} \tilde{r}_{a}^{1} -\frac{1}{2} \sum_{c, d} \hat{q}_{cd} q_{cd} - \sum_c \hat r_{c}^{2} r_{c}^{2}\\
		&= -\frac{1}{2}\left(\sum_a \hat{\tilde q}_{aa} \tilde{q}_{aa} + \sum_{a\neq b} \hat{\tilde q}_{ab} \tilde q_{ab} \right) -\frac{1}{2}\left(\sum_c \hat q_{cc} q_{cc} + \sum_{c\neq d} \hat q_{cd} q_{cd} \right)- n\hat{\tilde{r}}_1 \tilde{r}_1 - s\hat r_{2} r_{2}\\
		&= -\frac{1}{2}\left(n \hat{\tilde q}_{d} \tilde q_{d} + s(s-1) \hat{\tilde q}_{0} q_{0} \right) -\frac{1}{2}\left(s \hat q_{d} q_{d} + n(n-1) \hat q_{0} q_{0} \right) - n\hat{\tilde{r}}_1 \tilde{r}_1 - s\hat r_{2} r_{2}. \\
	\end{aligned}
\end{equation}
Then, the entropy term can be given by
\begin{equation}
	\begin{aligned}
		\mathcal{G}_{\mathrm{S}}
		&=  \int_{\tilde{\Omega}^{nN}} \int_{{\Omega}^{sN}} \prod_{a} \mathrm{d} \tilde{m}_{a} \prod_{c}\mathrm{d} m_{c} ~ e^{\frac{1}{2}\sum_{a,b} \hat{\tilde q}_{ab} \tilde m_a\tilde m_b +  +\frac{1}{2}\sum_{c,d} \hat{q}_{cd} m_c m_d-\beta\sum_c \mathrm{KL}(m_i^c, 	 \tilde m_i^{a=1}) }\\
		&\quad\times \left[ \frac{1+r_0}{2} \cosh\left( \sum_a \hat{\tilde{r}}_{a}^{1} \tilde{m}_i^a +\sum_c \hat{r}_{c}^{2} m_i^c  \right) +  \frac{1- r_0}{2} \cosh\left( \sum_a \hat{\tilde{r}}_{a}^{1} \tilde{m}_i^a -\sum_c \hat{r}_{c}^{2} m_i^c  \right)   \right] \\
		&=  \int_{\tilde{\Omega}^{ nN}} \int_{{\Omega}^{sN}} \prod_{a} \mathrm{d} \tilde{m}_{a} \prod_{c}\mathrm{d} m_{c}  \int \mathcal{D} z_1 \int \mathcal{D} z_2 e^{\frac{1}{2} (\hat{\tilde q}_{d} - \hat{\tilde q}_{0} )\sum_{a}(\tilde m_a)^2 +  \sqrt{\hat{\tilde q}_0} z_1\sum_a \tilde m_a + \frac{1}{2} (\hat{q}_{d} -  \hat{q}_{0}  )\sum_{c}(m_c)^2  +  \sqrt{\hat{q}_0} z_2 \sum_c  m_c}\\
		&\quad e^{-\beta\sum_c \mathrm{KL}(m_i^c, 	 \tilde m_i^{a=1})} \left[ \frac{1+r_0}{2} \cosh\left( \sum_a \hat{\tilde{r}}_{a}^{1} \tilde{m}_i^a +\sum_c \hat{r}_{c}^{2} m_i^c  \right) + \frac{1- r_0}{2} \cosh\left( \sum_a \hat{\tilde{r}}_{a}^{1} \tilde{m}_i^a -\sum_c \hat{r}_{c}^{2} m_i^c  \right)   \right]\\
		&= \frac{1+r_0}{2} \int \mathcal{D} z_1  {\left( \int_{-1}^{+1} \mathrm{d}\tilde{m}~ e^{\tilde{\mathcal{I}}(\tilde{m},z_1)}  \right)^{n-1}} {\int_{-1}^{+1} \mathrm{d}\tilde{m}~  e^{\tilde{\mathcal{I}}(\tilde{m},z_1)} \int \mathcal{D} z_2 \left(\int_{-1}^{+1} \mathrm{d}m ~e^{\mathcal{J}^{+}(m, \tilde{m}, z_2)} \right)^s}\\
		&\quad + \frac{1-r_0}{2} \int \mathcal{D} z_1 {\left( \int_{-1}^{+1} \mathrm{d}\tilde{m}~ e^{\tilde{\mathcal{I}}(\tilde{m},z_1)}  \right)^{n-1}} {\int_{-1}^{+1} \mathrm{d}\tilde{m}~  e^{\tilde{\mathcal{I}}(\tilde{m},z_1)}  \int \mathcal{D} z_2\left(\int_{-1}^{+1} \mathrm{d}m ~e^{\mathcal{J}^{-}(m, \tilde{m}, z_2)} \right)^s},
	\end{aligned}
\end{equation}
where
\begin{equation}
	\begin{aligned}
		\tilde{\mathcal{I}}(\tilde{m},z_1) &= \frac{1}{2} (\hat{\tilde q}_{d} - \hat{\tilde q}_{0} )\tilde m^2 + \left(\hat{\tilde{r}}_1 + \sqrt{\hat{\tilde q}_0} z_1\right) \tilde m,\\
		\mathcal{J}^{+}(m,\tilde{m},z_2) &= \frac{1}{2} (\hat{q}_{d} -  \hat{q}_{0})m^2 + \left(\hat{r}_{2} + \sqrt{\hat{q}_0} z_2 \right)  m  -\beta \mathrm{KL}(m, \tilde m),\\
		\mathcal{J}^{-}(m,\tilde{m},z_2) &= \frac{1}{2} (\hat{q}_{d} -  \hat{q}_{0})m^2  + \left(\hat{r}_{2} + \sqrt{\hat{q}_0} z_2 \right)  m  -\beta \mathrm{KL}(m, -\tilde m),\\
		{\rm KL}(x,y)&=-\sum_{z=\pm1}\left[\mathcal{K}(\frac{1+xz}{2},\frac{1+yz}{2})-\mathcal{K}(\frac{1+xz}{2},\frac{1+xz}{2})\right],
	\end{aligned}
\end{equation}
where $\mathcal{K}(x,y)=x\ln y$.
Finally, we derive the energy term, expressed as
\begin{equation}
	\begin{aligned}
		\mathcal{G}_{\mathrm{E}}^1
		&= \left\langle \prod_{a=1}^{n} H^{\tilde\beta}\left(-\frac{\operatorname{sign}({\tilde{v}_{1}}) \tilde u^{a}}{\sqrt{ 1-\tilde{q}_{aa}}}\right)\right\rangle\\
		&= \int \mathcal{D}z~2H\left(-\frac{\tilde{r}_{1}}{\sqrt{\tilde{q}_{0}-\tilde{r}_{1}^2}}z\right)\left( \int\mathcal{D}\sigma~H^{\tilde{\beta}}\left(-\frac{\sqrt{\tilde{q}_d-\tilde{q}_{0}} \sigma + \sqrt{\tilde{q}_{0}}z}{\sqrt{1-\tilde{q}_d}}\right)\right)^n,\\
	\end{aligned}
\end{equation}
and
\begin{equation}
	\begin{aligned}
		\mathcal{G}_{\mathrm{E}}^2
		&= \left\langle \prod_{c=1}^{s} H^{\beta}\left(-\frac{\operatorname{sign}({v_{2}}) u^{c}}{\sqrt{ 1-q_{cc}}}\right)\right\rangle\\
		&= \int \mathcal{D}z~2H\left(-\frac{{r_{2}}}{\sqrt{{q}_{0}-{r_{2}}^2}}z\right)\left( \int\mathcal{D}\sigma~H^{\beta}\left(-\frac{\sqrt{{q}_d-{q}_{0}} \sigma + \sqrt{{q}_{0}}z}{\sqrt{1-{q}_d}}\right)\right)^s,
	\end{aligned}
\end{equation}
where we follow the same computation as deriving Eq.~\eqref{Energy-term-t1}. Thus, we summarize the result as
\begin{equation}
	\mathcal{S} = -\frac{1}{2}\left(n \hat{\tilde q}_{d} \tilde q_{d} + s(s-1) \hat{\tilde q}_{0} q_{0} \right) -\frac{1}{2}\left(s \hat q_{d} q_{d} + n(n-1) \hat q_{0} q_{0} \right) - n\hat{\tilde{r}}_1 \tilde{r}_1 - s\hat r_{2} r_{2} + \ln \mathcal{G}_{\mathrm{S}} + \alpha_1 \ln \mathcal{G}_{\mathrm{E}}^1 + \alpha_2 \ln \mathcal{G}_{\mathrm{E}}^2.
\end{equation}

Calculation of saddle-point equations requires to consider the limits of $\lim_{n \rightarrow 0}$ and $\lim_{s \rightarrow 0}$, 
which leads to the computation of $\lim_{n \rightarrow 0}\lim_{s \rightarrow 0}\frac{\ln \mathcal{G}_{\mathrm{S}}}{n}$,
$\lim_{n \rightarrow 0}\lim_{s \rightarrow 0}\frac{\ln \mathcal{G}_{\mathrm{S}}}{s}$,
$\lim_{n \rightarrow 0}\lim_{s \rightarrow 0}\frac{\ln \mathcal{G}_{\mathrm{E}}^1}{n}$, and
$\lim_{n \rightarrow 0}\lim_{s \rightarrow 0}\frac{\ln \mathcal{G}_{\mathrm{E}}^2}{s}$. 
Thus, we define and compute these quantities first.
\begin{equation}
	\begin{aligned}
		\tilde{g}_{S} &=  \lim_{n \rightarrow 0}\lim_{s \rightarrow 0}\frac{\ln \mathcal{G}_{\mathrm{S}}}{n} \\
		&= \int\mathcal{D}z\ln \int_{-1}^{+1}  \mathrm{d} \tilde{m} ~e^{\tilde{\mathcal{I}}(\tilde{m},z)},\\
		g_{S} &= \lim_{n \rightarrow 0}\lim_{s \rightarrow 0}\frac{\ln \mathcal{G}_{\mathrm{S}}}{s} \\
		&=\frac{1+r_0}{2}  \int \mathcal{D} z_1 \left\langle \left\langle \int \mathcal{D} z_2 \ln \int_{-1}^{+1} e^{\mathcal{J}^{+}(m,\tilde{m},z_2)}  \right\rangle \right\rangle_{\tilde{\mathcal{I}}(\tilde{m},z_1)}\\
		& \quad + \frac{1-r_0}{2}  \int \mathcal{D} z_1 \left\langle \left\langle \int \mathcal{D} z_2 \ln \int_{-1}^{+1} e^{\mathcal{J}^{-}(m,\tilde{m},z_2)}  \right\rangle \right\rangle_{\tilde{\mathcal{I}}(\tilde{m},z_1)},\\
		\tilde{g}_E &= \lim_{n \rightarrow 0}\lim_{s \rightarrow 0}\frac{\ln \mathcal{G}_{\mathrm{E}}^1}{n}\\
		&=\int\mathcal{D}z~2H\left(-\frac{\tilde{r}_{1}}{\sqrt{\tilde{q}_{0}-\tilde{r}_{1}^2}}z\right)\ln\int\mathcal{D}\sigma~H^{\tilde{\beta}}\left(-\frac{\sqrt{\tilde{q}_d-\tilde{q}_{0}} \sigma + \sqrt{\tilde{q}_{0}}z}{\sqrt{1-\tilde{q}_d}}\right),\\
		g_E &= \lim_{n \rightarrow 0}\lim_{s \rightarrow 0}\frac{\ln \mathcal{G}_{\mathrm{E}}^2}{s}\\
		& = \int \mathcal{D}z~2H\left(-\frac{r_{2}}{\sqrt{q_0-r_{2}^2}}z\right)\ln \int\mathcal{D}\sigma~H^{\beta}\left(-\frac{\sqrt{q_d-q_0} \sigma + \sqrt{q_0}z}{\sqrt{1-q_d}}\right).
	\end{aligned}
\end{equation}
Then, we can arrive at the saddle-point equations given below.
\begin{equation}
	\begin{aligned}
		\tilde{q}_{d} &= 2\frac{\partial \tilde{g}_{S} }{\partial \hat{\tilde{q}}_{d}},&
		\tilde{q}_{0} &= -2\frac{\partial \tilde{g}_{S}}{\partial \hat{\tilde{q}}_{0}},&
		\tilde{r}_{1} &= \frac{\partial \tilde{g}_{S}}{\partial \hat{\tilde{r}}_{1}},&
		\hat{\tilde{q}}_{d} &= 2\alpha_1\frac{\partial \tilde{g}_{E}}{\partial \tilde{q}_{d}},&  \hat{\tilde{q}}_{0} &= -2\alpha_1\frac{\partial \tilde{g}_{E}}{\partial \tilde{q}_{0}},&
		\hat{\tilde{r}}_{1} &= \alpha_1\frac{\partial \tilde{g}_{E}}{\partial \tilde{r}_{1}};\\
		q_{d} &= 2\frac{\partial g_{S} }{\partial \hat{q}_{d}},&
		q_{0} &= -2\frac{\partial g_{S}}{\partial \hat{q}_{0}},&
		r_{2} &= \frac{\partial g_{S}}{\partial \hat{r}_{2}},&
		\hat{q}_{d} &= 2\alpha_2\frac{\partial g_{E}}{\partial q_{d}},&
		\hat{q}_{0} &= -2\alpha_2\frac{\partial g_{E}}{\partial q_{0}},&
		\hat{r}_{2} &= \alpha_2\frac{\partial g_{E}}{\partial r_{2}}.
	\end{aligned}
\end{equation}

It is easy to verify that the tilde order parameters
are exactly the same as those in Eq.~(\ref{sd-t1}), which are independent of
the non-tilded order parameters. This is reasonable because
in the multi-task loss function $\mathcal{L}_2(\boldsymbol{m},\tilde{\boldsymbol{m}})$, the
magnetization $\tilde{\boldsymbol{m}}$ in the KL-divergence is the solution
after learning the first task, which is described by the single-task 
partition function Eq.~(\ref{Z-t1}). As for the non-tilded order parameters, 
the hatted ones are in the same form with Eqs.~(\ref{sd-task1-ge},\ref{sd-task1-ge-where}),
except for the replacement of $r_1$ by $r_2$ in Eq.~(\ref{sd-task1-ge-where}). 
After a few manipulations, the second-task related order parameters are expressed as follows,
\begin{equation}
	\begin{aligned}
		q_{d} &= \frac{1+r_0}{2}  \int \mathcal{D} z_1 \left\langle \left\langle \int \mathcal{D} z_2  \left\langle\left\langle m^2 \right\rangle \right\rangle_{\mathcal{J}^{+}(m,\tilde{m},z_2)} \right\rangle \right\rangle_{\tilde{\mathcal{I}}(\tilde{m},z_1)}\\
		&\quad + \frac{1-r_0}{2}  \int \mathcal{D} z_1 \left\langle \left\langle \int \mathcal{D} z_2  \left\langle\left\langle m^2 \right\rangle \right\rangle_{\mathcal{J}^{-}(m,\tilde{m},z_2)} \right\rangle \right\rangle_{\tilde{\mathcal{I}}(\tilde{m},z_1)},\\
		q_{0} &=\frac{1+r_0}{2}  \int \mathcal{D} z_1 \left\langle \left\langle \int \mathcal{D} z_2  \left\langle\left\langle m^2 - \frac{z}{\sqrt{q_0}}m \right\rangle \right\rangle_{\mathcal{J}^{+}(m,\tilde{m},z_2)} \right\rangle \right\rangle_{\tilde{\mathcal{I}}(\tilde{m},z_1)}\\
		&\quad + \frac{1-r_0}{2}  \int \mathcal{D} z_1 \left\langle \left\langle \int \mathcal{D} z_2  \left\langle\left\langle m^2 - \frac{z}{\sqrt{q_0}}m \right\rangle \right\rangle_{\mathcal{J}^{-}(m,\tilde{m},z_2)} \right\rangle \right\rangle_{\tilde{\mathcal{I}}(\tilde{m},z_1)},\\
		r_{2} &= \frac{1+r_0}{2} \int \mathcal{D} z_1 \left\langle \left\langle \int \mathcal{D} z_2  \left\langle\left\langle m \right\rangle \right\rangle_{\mathcal{J}^{+}(m,\tilde{m},z_2)} \right\rangle \right\rangle_{\tilde{\mathcal{I}}(\tilde{m},z_1)}\\
		&\quad + \frac{1-r_0}{2} \int \mathcal{D} z_1 \left\langle \left\langle \int \mathcal{D} z_2  \left\langle\left\langle m \right\rangle \right\rangle_{\mathcal{J}^{-}(m,\tilde{m},z_2)} \right\rangle \right\rangle_{\tilde{\mathcal{I}}(\tilde{m},z_1)}.\\
	\end{aligned}
\end{equation}

\subsubsection{Generalization error for two tasks}
The derivation of generalization error in the multi-task scenario follows
the same procedure with the single-task scenario. Thus, we present the final results directly.
After the convergence of all order parameters, the generalization error for task 2 reads
\begin{equation}
	\epsilon_{g}^{2} =  \frac{1}{\pi}\arccos (p_2),
\end{equation}
where
\begin{equation}
	\begin{aligned}
		p_{2} &=\frac{1+r_0}{2} \int \mathcal{D} z_1 \left\langle \left\langle \int \mathcal{D} z_2  \left\langle\left\langle \operatorname{sign}(m) \right\rangle \right\rangle_{\mathcal{J}^{+}(m,\tilde{m},z_2)} \right\rangle \right\rangle_{\tilde{\mathcal{I}}(\tilde{m},z_1)}\\
		&\quad + \frac{1-r_0}{2} \int \mathcal{D} z_1 \left\langle \left\langle \int \mathcal{D} z_2  \left\langle\left\langle \operatorname{sign}(m) \right\rangle \right\rangle_{\mathcal{J}^{-}(m,\tilde{m},z_2)} \right\rangle \right\rangle_{\tilde{\mathcal{I}}(\tilde{m},z_1)}.\\
	\end{aligned}
\end{equation}
The generalization error for task 1 reads
\begin{equation}
	\epsilon_{g}^{1} =  \frac{1}{\pi}\arccos (p_1),
\end{equation}
where
\begin{equation}
	\begin{aligned}
		p_{1} &= \frac{1+r_0}{2} \int \mathcal{D} z_1 \left\langle \left\langle \int \mathcal{D} z_2  \left\langle\left\langle \operatorname{sign}(m) \right\rangle \right\rangle_{\mathcal{J}^{+}(m,\tilde{m},z_2)} \right\rangle \right\rangle_{\tilde{\mathcal{I}}(\tilde{m},z_1)}\\
		& \quad - \frac{1-r_0}{2} \int \mathcal{D} z_1 \left\langle \left\langle \int \mathcal{D} z_2  \left\langle\left\langle \operatorname{sign}(m) \right\rangle \right\rangle_{\mathcal{J}^{-}(m,\tilde{m},z_2)} \right\rangle \right\rangle_{\tilde{\mathcal{I}}(\tilde{m},z_1)}.\\
	\end{aligned}
\end{equation}
\subsubsection{The case of tunned $\mathrm{KL}$ terms}
To investigate the regularization term, we can multiply this term with a factor $\gamma$, and then derive the saddle point equations for the multi-task learning as above.
Finally we can change the value of the modulation factor to probe effects of the regularization term. The objective function then reads
\begin{equation}
	\mathcal{L}_2(\boldsymbol{m},\tilde{\boldsymbol{m}},\gamma)=\sum_{\mu=1}^{M_2} \ln H\left(-\frac{\operatorname{sign}(\sum_{i} W_i^{1}x_i^{2,\mu})\sum_{i} m_{i} x_i^{2,\mu}}{\sqrt{\sum_{i}\left(1-m_{i}^{2}\right)}}\right) - \gamma\sum_{i=1}^N \mathrm{KL}(Q_{m_i}|Q_{m_i^{1}}).
\end{equation}
This minor change will not affect the whole calculation process, but only induce a corresponding factor in the auxiliary terms,
\begin{equation}
	\begin{aligned}
		\mathcal{J}^{+}(m,\tilde{m},z_2,\gamma) &= \frac{1}{2} (\hat{q}_{d} -  \hat{q}_{0})m^2 + \left(\hat{r}_{2} + \sqrt{\hat{q}_0} z_2 \right)  m  -\gamma \beta \mathrm{KL}(m, \tilde m),\\
		\mathcal{J}^{-}(m,\tilde{m},z_2,\gamma) &= \frac{1}{2} (\hat{q}_{d} -  \hat{q}_{0})m^2  + \left(\hat{r}_{2} + \sqrt{\hat{q}_0} z_2 \right)  m  -\gamma \beta \mathrm{KL}(m, -\tilde m).
	\end{aligned}
\end{equation}
Thus, the saddle-points equations remain the same except for the following differences,
\begin{equation}
	\begin{aligned}
		q_{d} &= \frac{1+r_0}{2}  \int \mathcal{D} z_1 \left\langle \left\langle \int \mathcal{D} z_2  \left\langle\left\langle m^2 \right\rangle \right\rangle_{\mathcal{J}^{+}(m,\tilde{m},z_2,\gamma)} \right\rangle \right\rangle_{\tilde{\mathcal{I}}(\tilde{m},z_1)}\\
		&\quad + \frac{1-r_0}{2}  \int \mathcal{D} z_1 \left\langle \left\langle \int \mathcal{D} z_2  \left\langle\left\langle m^2 \right\rangle \right\rangle_{\mathcal{J}^{-}(m,\tilde{m},z_2,\gamma)} \right\rangle \right\rangle_{\tilde{\mathcal{I}}(\tilde{m},z_1)},\\
		q_{0} &=\frac{1+r_0}{2}  \int \mathcal{D} z_1 \left\langle \left\langle \int \mathcal{D} z_2  \left\langle\left\langle m^2 - \frac{z}{\sqrt{q_0}}m \right\rangle \right\rangle_{\mathcal{J}^{+}(m,\tilde{m},z_2,\gamma)} \right\rangle \right\rangle_{\tilde{\mathcal{I}}(\tilde{m},z_1)}\\
		&\quad + \frac{1-r_0}{2}  \int \mathcal{D} z_1 \left\langle \left\langle \int \mathcal{D} z_2  \left\langle\left\langle m^2 - \frac{z}{\sqrt{q_0}}m \right\rangle \right\rangle_{\mathcal{J}^{-}(m,\tilde{m},z_2,\gamma)} \right\rangle \right\rangle_{\tilde{\mathcal{I}}(\tilde{m},z_1)},\\
		r_{2} &= \frac{1+r_0}{2} \int \mathcal{D} z_1 \left\langle \left\langle \int \mathcal{D} z_2  \left\langle\left\langle m \right\rangle \right\rangle_{\mathcal{J}^{+}(m,\tilde{m},z_2,\gamma)} \right\rangle \right\rangle_{\tilde{\mathcal{I}}(\tilde{m},z_1)}\\
		&\quad + \frac{1-r_0}{2} \int \mathcal{D} z_1 \left\langle \left\langle \int \mathcal{D} z_2  \left\langle\left\langle m \right\rangle \right\rangle_{\mathcal{J}^{-}(m,\tilde{m},z_2,\gamma)} \right\rangle \right\rangle_{\tilde{\mathcal{I}}(\tilde{m},z_1)}.\\
	\end{aligned}
\end{equation}


\begin{acknowledgments}
We thank the referee for many constructive comments to improve the quality of the paper.
This research was supported by
the National Key R$\&$D Program of China
(2019YFA0706302) and the National Natural Science Foundation of China for
Grant Number 12122515 (H.H.), and the National Natural Science Foundation of China for Grant Number 11975295 (Z.H.), and Guangdong Provincial Key Laboratory of Magnetoelectric Physics and Devices (No. 2022B1212010008), 
and Guangdong Basic and Applied Basic Research Foundation (Grant No. 2023B1515040023).  
\end{acknowledgments}


\end{document}